\documentclass[10pt,preprint,flushrt]{aastex}
\usepackage{natbib}
\usepackage{graphics}
\usepackage{amssymb,amsmath}
\usepackage{graphicx}
\usepackage{mathrsfs}
\usepackage{subfigure}
\newcommand{\refeqn}[1]{Eq. (\ref{#1})} \newcommand{\reffig}[1]{Fig. \ref{#1}}
\newcommand{\ld}{\lambda/D}
\newcommand{\xinaught}{\xi_{\mbox{\scriptsize iwa}}}
\newcommand{\xione}{\xi_{\mbox{\scriptsize owa}}}

\bibdata{FinalReport.bib}

\begin{abstract}
The realization that direct imaging of extrasolar planets could be technologically feasible within the next decade or so has inspired a great deal of recent research into high-contrast imaging.  We ourselves have contributed several design ideas, all of which can be described as {\em shaped pupil coronagraphs}.  In this paper, we offer a complete and unified survey of asymmetric shaped pupils designs, some of which have been published in our previous papers.  We also introduce a promising new design, which we call {\em barcode masks}.  These masks achieve the required contrast with a fairly large discovery zone and throughput but most importantly they are perhaps the easiest to manufacture and might therefore stand up best to a refined analysis based
on vector propogation techniques.
\end{abstract}

\begin{document}
\title{Optimal Asymmetric Apodizations and Shaped Pupils for Planet Finding Coronagraphy}
\author{ N. Jeremy Kasdin, Robert J. Vanderbei, Michael G. Littman, David N. Spergel}
\affil{Princeton University \\ Princeton, NJ   08544}

\maketitle

\section{Introduction}

The discovery of more than 100 extrasolar Jupiter-sized planets in just the last decade has generated enormous interest, both among astronomers and the public, in the problem of discovering and characterizing Earthlike planets.  NASA is already planning its next large space-based observatory, the \emph{Terrestrial Planet Finder (TPF)}, with a planned launch date toward the end of the next decade.  TPF's primary objective will be to discover Earthlike planets and characterize them for indications of life.  

While the technical challenges for TPF are great,  foremost among them is the problem of high-contrast imaging.  In order to discover as many planets as possible, it is necessary to design an imaging system that achieves very high contrast between the parent star and the nearby planet.  An earlier study by \cite{ref:Brown} 
indicates that a $D = 4$m class visible-light 
(i.e. $400\text{nm} \le \lambda \le 650\text{nm}$)
instrument ought to be able to discover
about 50 extrasolar Earth-like planets if it can provide contrast
of $10^{-10}$ at an angular separation of $3\ld$ and that a $4 \times 10$m class
telescope ought to be able to discover about 150 such
planets if it can provide the same contrast at a separation of $4\ld$.

Many approaches have been examined for designing a telescope with this level of contrast.  The most promising fall into two broad categories---nulling interferometers (operating in the infrared) and coronagraphs (operating in the visible).  One important subset of coronagraphs, referred to as shaped pupils, has been gaining interest (\cite{ref:Spergel, ref:KVSL, VSK02, VSK03}).  These are apodized entrance pupils that rely solely on one/zero binary openings.   In this paper we present a unified treatment of the one-dimensional shaped pupil designs, a few of which have been previously published, and we introduce a promising new design which we call {\em barcode masks}.


The paper is organized as follows.  In the next section we briefly review the relationship between pupil-plane apodization 
and the corresponding image-plane electric field and point-spread function.  In the following section we discuss the various performance metrics that characterize a planet finding telescope.  In Section 4 we show how optimization problems can be formulated and used to find the ``best'' entrance pupil apodization for high contrast in one-dimensional square apertures as well as azimuthally symmetric (circular or elliptical) pupils.  In Section 5 we describe how these results can be used to find various families of assymmetric shaped pupils.  In section 6 we present a preliminary sensitivity analysis to determine the tolerances of the pupil approaches to manufacturing errors.

\section{Apodization}

The main objective of our work is to develop optimal, high-contrast, binary shaped pupils (that is, pupils that consist of only open and closed areas).  Since pupil masks can be viewed mathematically as a special case of apodized pupils (where the apodization function is simply zero-one valued), we begin by developing some notation and concepts in the general context of pupil apodization.  In this paper we use scalar far-field diffraction theory (i.e., Fraunhoffer optics).  We are currently investigating near field and vector propagation effects and will report on those more refined models in a future paper.

In the Fraunhoffer regime, the image-plane electric field $E(\xi,\zeta)$, produced by an on-axis point source and an \emph{apodization function}, $A(x,y)$, reduces to the Fourier Transform of the apodization:
\begin{equation}
\label{eq:squareapod}
E(\xi,\zeta)=\int{\int_{S}{e^{-2\pi i (x\xi+y\zeta)}A(x,y)dxdy}}
\end{equation}
where S denotes the overall aperture and can be taken, for example, to be
a square:
\[ S=\{(x,y):-1/2\le x \le 1/2, -1/2\le y \le 1/2\}  \]
 
 Here, $x$ and $y$ denote the coordinates in the pupil plane measured in units of the aperture $D$ and ($\xi$, $\zeta$) denote angular (radian) deviations from on-axis measured in units of wavelength over aperture ($\lambda/D$) or, equivalently, physical distance in the image plane measured in units of focal-length times wavelength over aperture ($f\lambda/D$).  

When S is a square and the apodization function, $A(x)$,  is one-dimensional and symmetric about the $y$-axis, the expression for the electric field can be simplified into a single, one-dimensional Fourier Transform:
\begin{equation}
\label{eq:onedapod}
E(\xi,\zeta)=\frac{2\sin(\pi\zeta)}{\pi\zeta}\int_0^{1/2}{A(x)\cos(2\pi \xi x) dx }
\end{equation}

In this paper we will only be considering apodizations (and shaped pupils)  along a single dimension. 
    For both square and circular  apertures, the {\em point spread function} (PSF) is the square of the electric field and, for planet finding, the apodization is chosen so that the corresponding PSF has the desired contrast.

\section{Performance Metrics}

The goal of our apodizations is to create point spread functions that provide the contrast necessary to detect a planet \emph{with as short an integration time as possible}.  In order to compare different apodization approaches and to perform optimizations it is therefore necessary to introduce a few important performance metrics (\cite{VSK03}).  The first is {\em contrast}, which is a function of the linear  position in the image plane.  In rectangular coordinates we define contrast by:
        \[
    E^2(\xi,\zeta)/E^2(0,0).
    \]
    %
    We say that we have high-contrast when this ratio is very
    small.  For planet finding, the consensus is that a contrast of $10^{-10}$ outside a predefined radius is
    necessary in order to image an Earth-like planet that is $1$ AU from its
    Sun-like star.

 The second and third metrics are the {\em inner working angle}, $\xinaught$,
    and, of lesser importance, the {\em outer working angle}, $\xione$.
    We wish to design pupil masks that have high-contrast for all $\xi$'s
    in the interval $\xinaught \le \xi \le \xione$.  
    
The final performance metric is integration time, or its surrogate, throughput.  Integration time is a critical metric for any planet finding telescope such as TPF.  Smaller integration time relaxes requirements on system stability, simplifies control, allows more observations, and potentially allows return visits to interesting systems.  In ~\cite{ref:KVSL}  we described a number of different approaches to image analysis that resulted in quantifiable integration times that we used to compare different coronagraph approaches.  All are based on some type of photometry on a potential planet and a predefined signal-to-noise ratio.  While useful for comparing different coronagraphs, the resulting equations were too complicated to use in optimization work.  Therefore, we consider here slightly simpler measures of throughput. (These measures are closely related to integration time because they measure the amount of light entering the system, but they do not account for the effect of PSF shape.  While we acknowledge that PSF shape or ``sharpness'' is an important quantity in determining absolute integration times, it does not have a strong effect on the relative comparisons or optimizations.)  Even asking only for measures of throughput is not such a simple question as there are multiple ways one might formulate an answer.

Perhaps the most natural measure of throughput is the amount of energy that falls into the main lobe of the PSF relative to the total energy conveyed through a fully open aperture:
\begin{equation}
\mathcal{T}_{Airy} =4 \int_0^{\xi_{iwa}}{\int_0^{\zeta_{iwa}}{E^2(\xi,\zeta) d\zeta d\xi}}
\end{equation}
where $\xi_{\mbox{iwa}}$ and $\zeta_{\mbox{iwa}}$ denote the locations in ($\xi,\zeta$) of the first null of the PSF.  We call this the \emph{Airy-throughput}  (\cite{VSK03}).  

There are two other relevant measures of throughput.  The {\em total throughput} of an apodization is the integral of the PSF over the entire image plane, which,  by Parseval's theorem, is the integral of the square of the apodization over the entire entrance pupil.

Finally, the electric field at ($\xi=0, \zeta=0$),
\begin{equation}
E(0,0) = 2\int_0^{1/2}{A(x)dx}
\end{equation}
for one-dimensional rectangular apodizations provides another measure of the ``central'' throughput of the apodization, since its square is the peak throughput density at the center of the Airy disk.  If $A()$ is zero-one valued, then the apodization can be realized as a shaped pupil mask.  In this case, $E(0,0)$ is precisely the open area of the mask (and is also the total-throughput).  For this reason, we call $E(0,0)$ the {\em pseudo-area} of the apodization, even when the apodization is not zero-one valued.

\section{Optimal Apodizations for High-Contrast}
\label{sec:optimal_apod}

In this section we discuss the general problem of designing apodized pupils for high contrast.  This problem is not new; an excellent survery can be found in  \citet{ref:Jacquinot} with more recent ideas in \cite{ref:Indebetouw} and \cite{ref:Watson}.   \cite{ref:Nisenson} describe their concept of using an apodized square aperture for high-contrast, where discovery is made along the diagonals.  We describe below alternative optimal apodizations that  lead naturally to various shaped pupil concepts.

\subsection{One-Dimensional Apodizations}
\label{sec:one_d}

\begin{figure}[t]
\plottwo{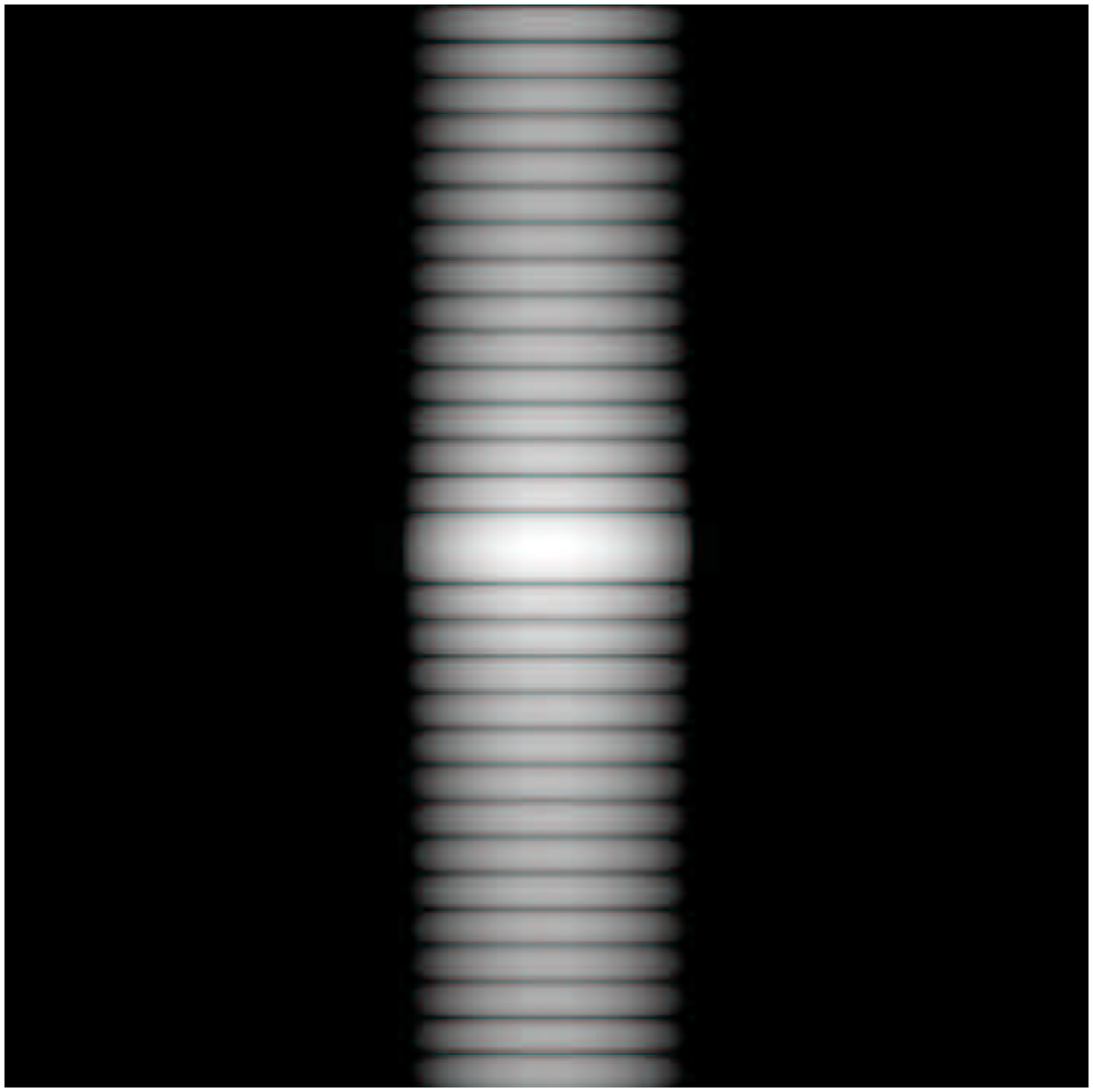}{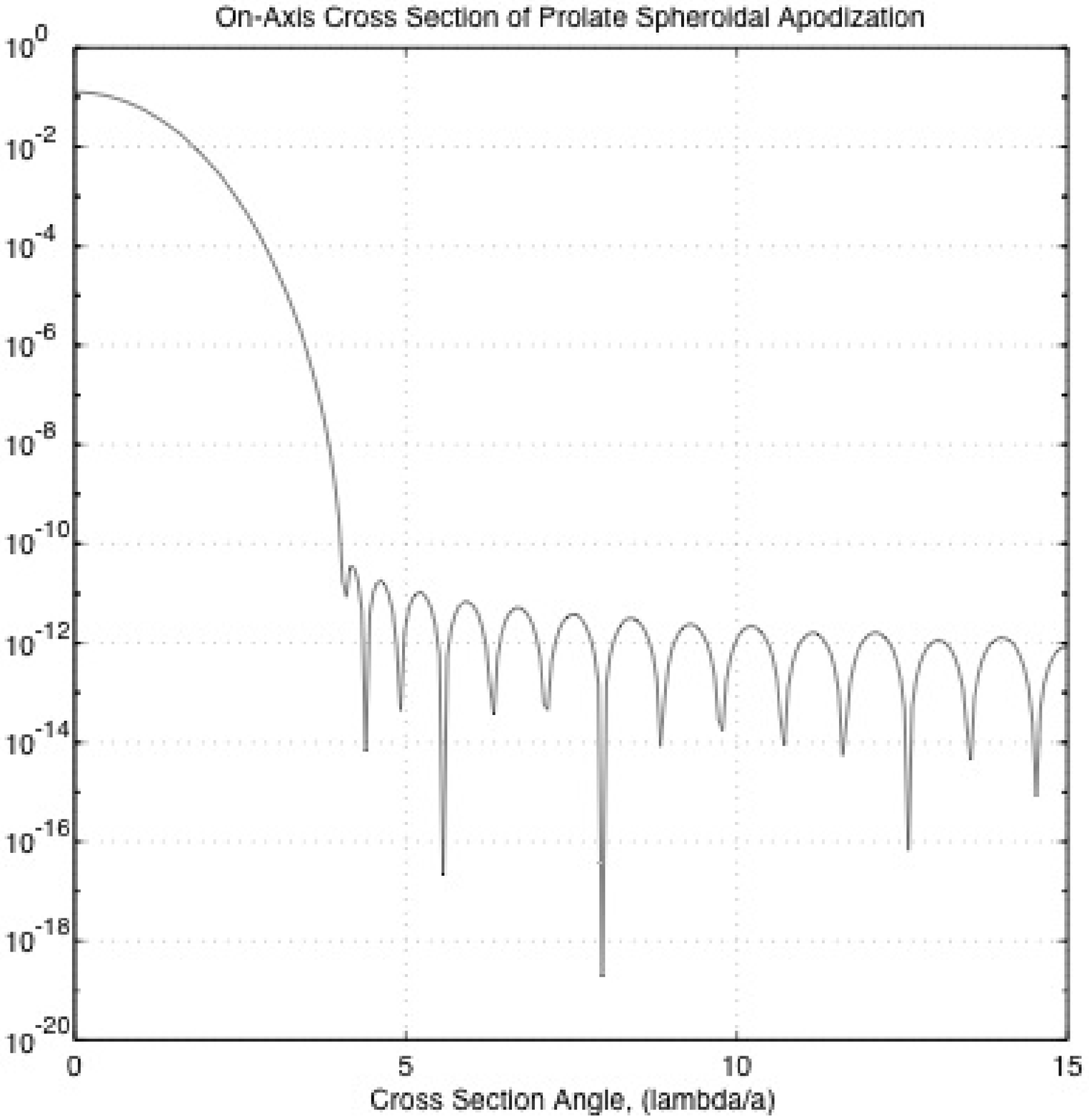}
\begin{center}
\end{center}
\caption{{\em Left} The PSF for the one-dimensional prolate spheroidal apodized square
aperture~\citep{ref:Slepian} of unit area plotted
on a logarithmic scale with black areas $10^{-10}$ below brightest.
{\em Right} On-axis cross section of the PSF showing an inner working distance of
$4 \lambda/a$.  The total throughput is 25\%.}
\label{fig:OneDProlate}
\end{figure}

Based on \refeqn{eq:onedapod} we ask for the optimal one-dimensional apodization for planet finding.  Of course, there are a number of ways one can state the optimization problem, all involving shaping the point spread function (PSF) in some way.  For instance, \cite{ref:Jacquinot} describe many different approaches to designing optimal apodizations.  The most important optimal apodization is due to \cite{ref:Slepian}, who introduced the prolate spheroidal wave function as the optimal 1-D telescope apodization.  Slepian's one dimensional optimization problem  asks for the function that concentrates as much light as possible into the central lobe of the finite Fourier transform.  \cite{ref:Slepian1} derive a finite fourier transform analog of the uncertainty principal and show that the function that solves this optimization problem is the zero order prolate spheroidal wavefunction, that is, the solution to the wave equation in prolate spheroidal coordinates (there is also an elegant solution to this problem using the calculus of variations).   Their statement is equivalent to the computationally easier problem:
 \[
        \begin{array}{ll}
            \mbox{minimize } & 
	    \displaystyle \int_{\xi_{iwa}}^{\infty} E(\xi,0)^2 d\xi \\[0.2in]
	    \mbox{subject to } &
	        \setlength{\arraycolsep}{0.1em}
	        \begin{array}[t]{rcll}
	        \displaystyle
		    & A(0) & = 1 .
	        \end{array}
        \end{array}
    \]
Choosing $\xinaught = 4$ provides the needed $10^{-10}$ contrast for $\xi \ge \xinaught$.  \reffig{fig:OneDProlate} shows the PSF resulting from a square aperture apodized in one dimension with a prolate spheroidal wave function.  The total  throughput of this apodized system is 25\%.

\begin{figure}[t]
\begin{center}
\includegraphics[width=2 in]{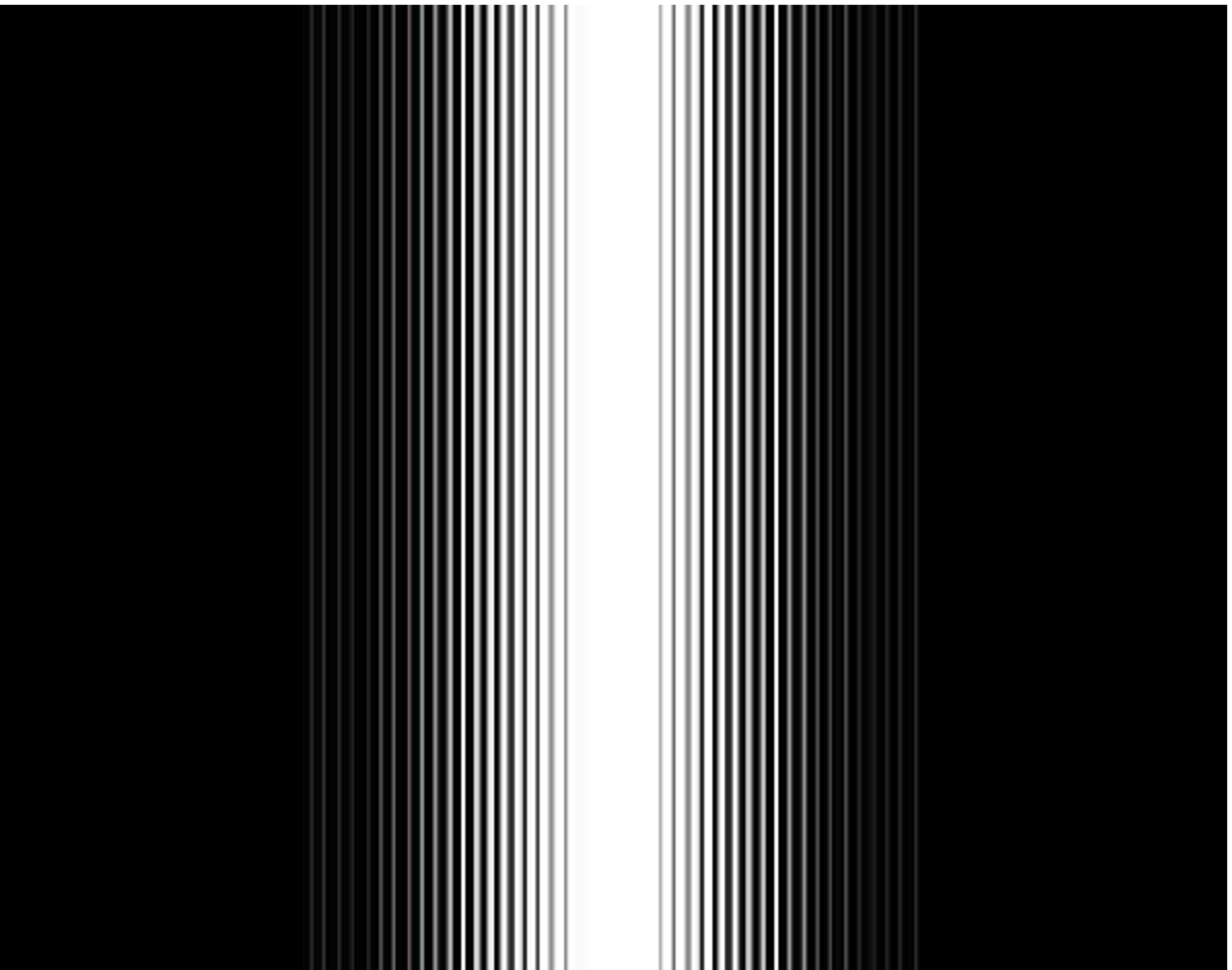}
\includegraphics[width=2 in]{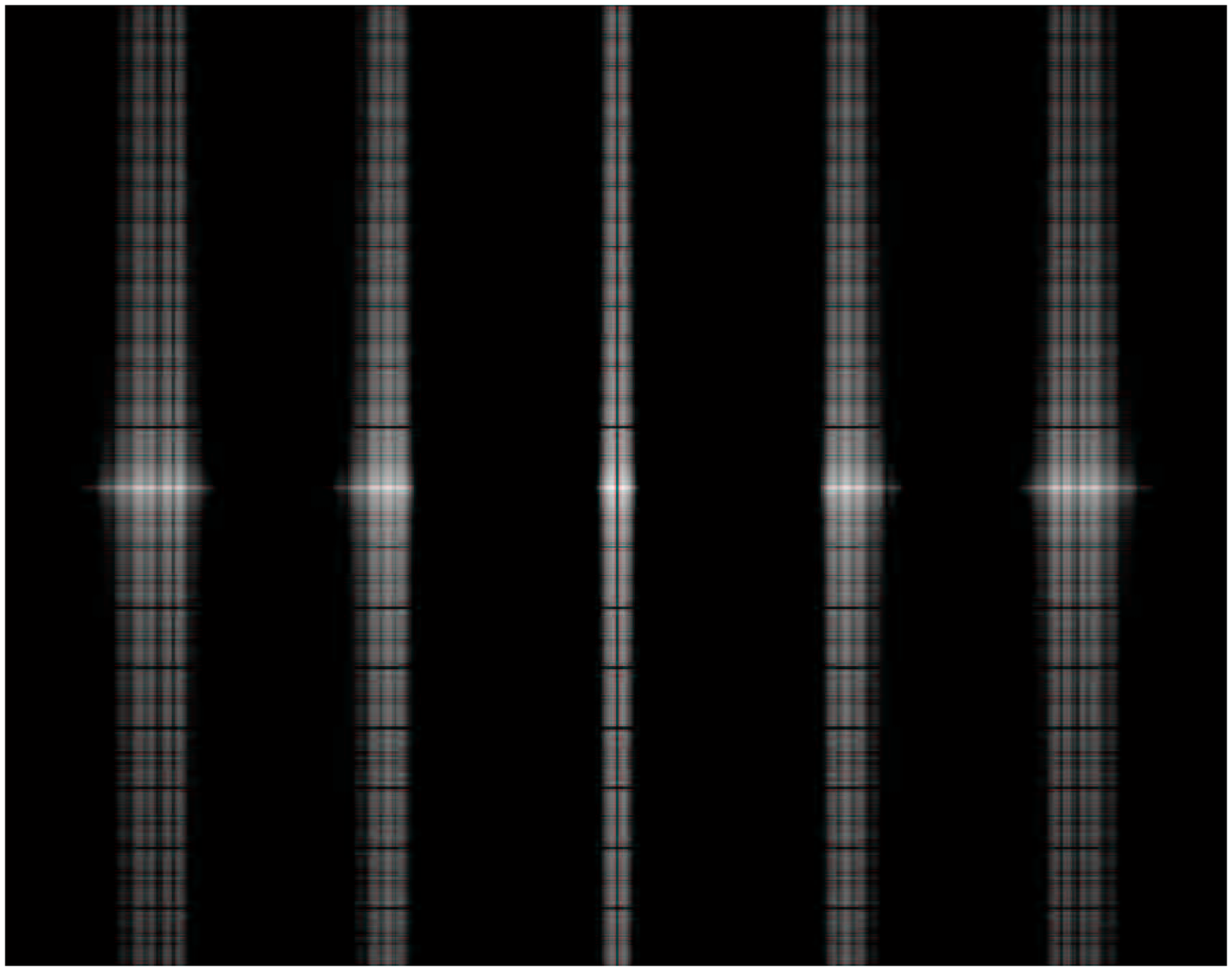}
\includegraphics[width=2 in, angle=90, origin=c]{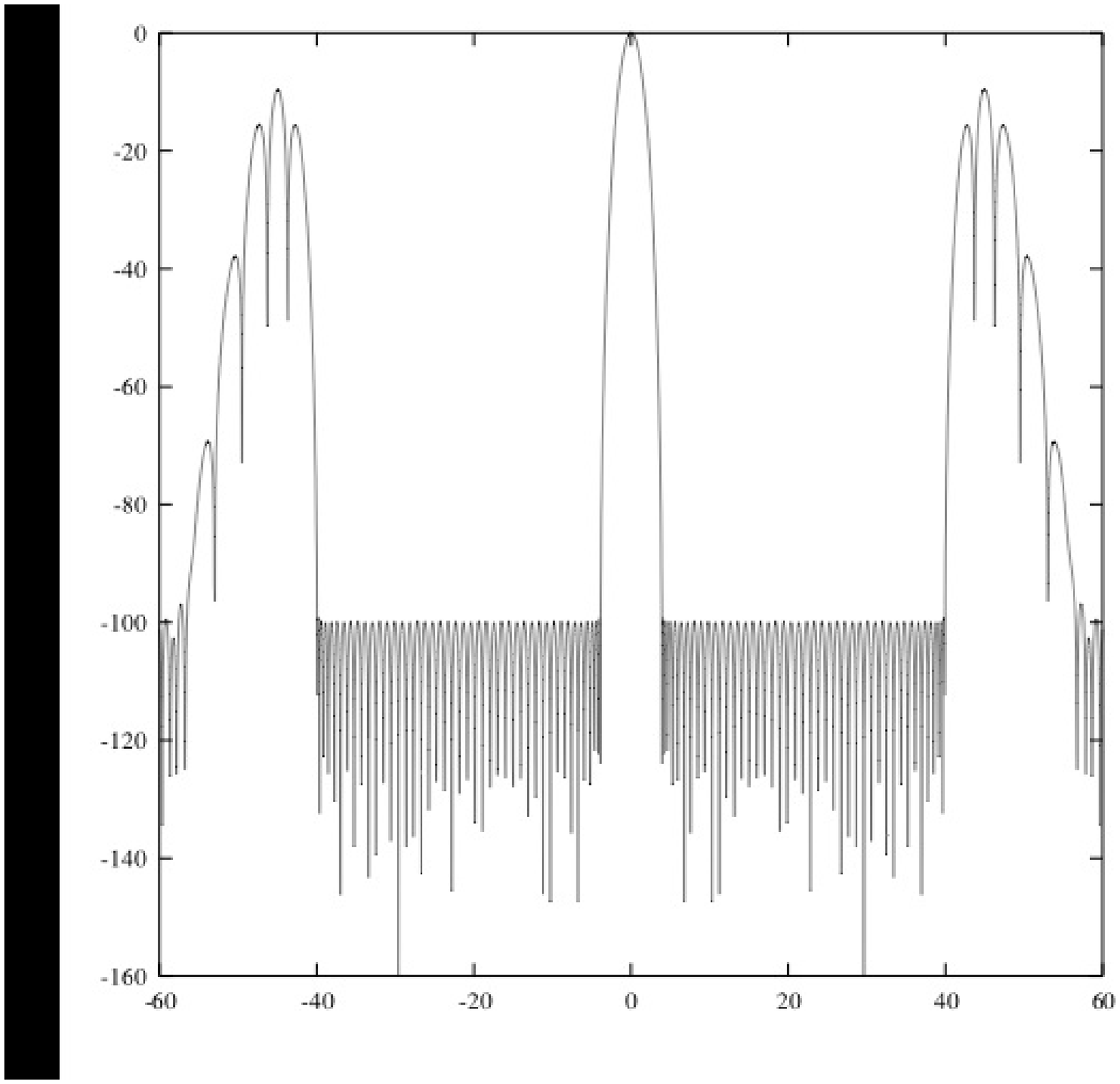}
\end{center}
\caption{{\em Left} A one-dimensional barcode mask optimized for a dark region in [0,40] (all slots are not visible because of resolution limitations). {\em Center} The 2-D point spread function of the barcode mask {\em Right} A cross section of the optimal one-dimensional PSF near the inner working angle.  The Airy throughput is 25 \%.} 
\label{fig:barcode}
\end{figure}

We note that outside the iwa the PSF has higher than needed contrast, which leads us to a slightly different optimization problem---namely, to maximize the pseudo-area subject to contrast constraints.  This has the enormous advantage that it can be formulated as an infinite dimensional linear programming problem:

\begin{equation} \label{100}
    \begin{array}{ll}
        \mbox{maximize } & \int_0^{1/2} A(x) dx \\
	\mbox{subject to } &
	    \setlength{\arraycolsep}{0.1em}
	    \begin{array}[t]{rcll}
	    \displaystyle
	        -10^{-5} E(0,\zeta) \le & E(0,\zeta) & \le 10^{-5} E(0,\zeta), &
	            \qquad \xinaught \le \xi \le \xione , \\
		  0 \le & A(x)   & \le  1, & \qquad 0 \le x \le 1/2, \\
	    \end{array}
    \end{array}
\end{equation}
Discretizing the sets of $x$'s and $\xi$'s and replacing the integrals with their Riemann sums, problem~(\ref{100}) is approximated by a finite dimensional linear programming problem, which can be solved to a high level of precision (see, e.g., \citet{Van01}).
The numerical solution to this problem reveals that the optimal solution is
zero-one valued.   In other words, the optimal solution is, in fact, a binary mask consisting of a series of slots in the $y$-direction of varying width alternating along the $x$-axis.  We call such a mask a {\em barcode} mask.  \reffig{fig:barcode} shows a cross section of the PSF of such a mask optimized for high contrast in the region $4 \le \xi \le 40 \mbox{ } (\lambda/D)$.  The Airy throughput of this mask is 25\%.  Note that this response is very similar to the sine-squared bandlimited coronagraph, with bands of discovery zones in the image plane (\cite{ref:kuchner}).  

\begin{figure}[t]
\begin{center}
\plottwo{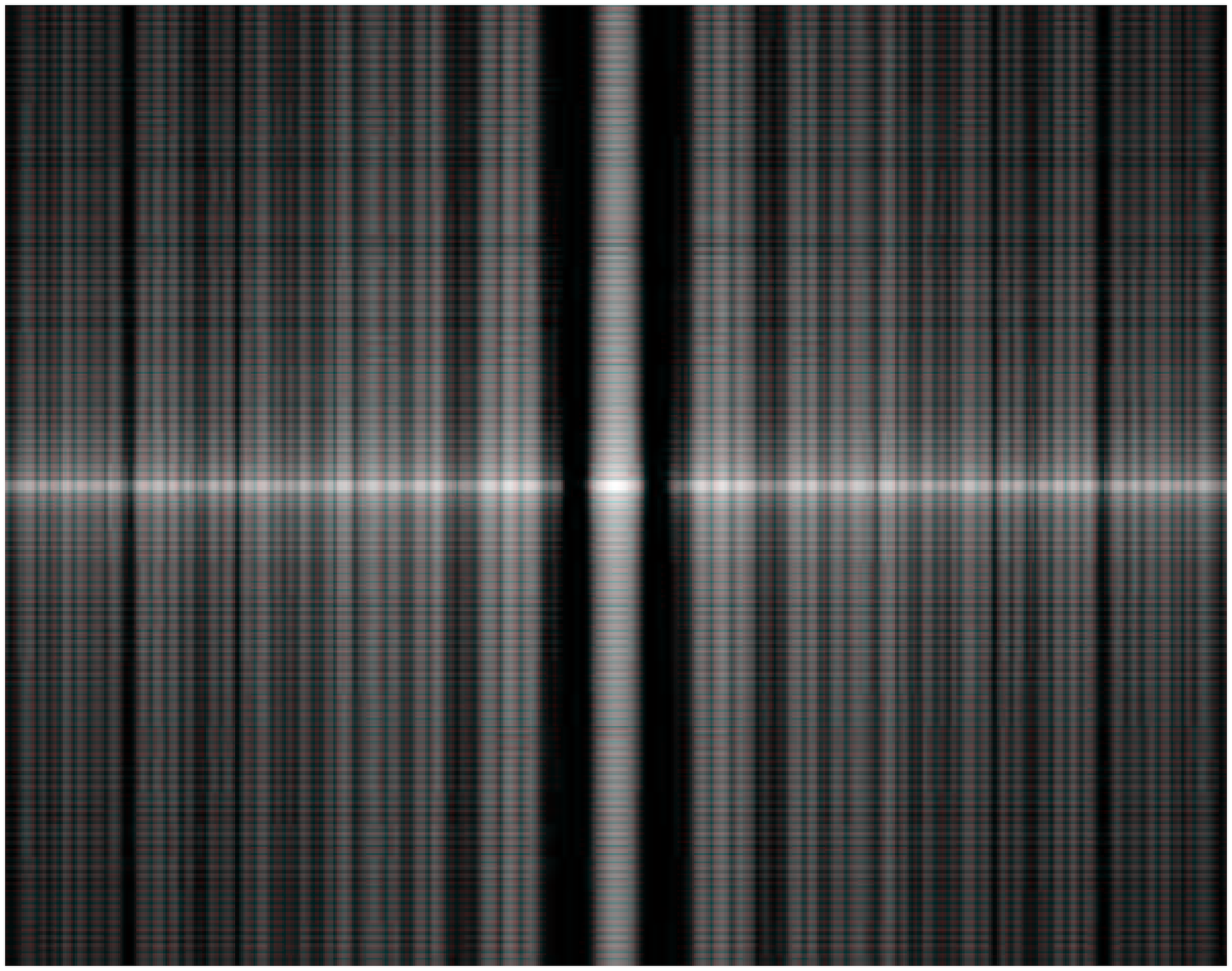}{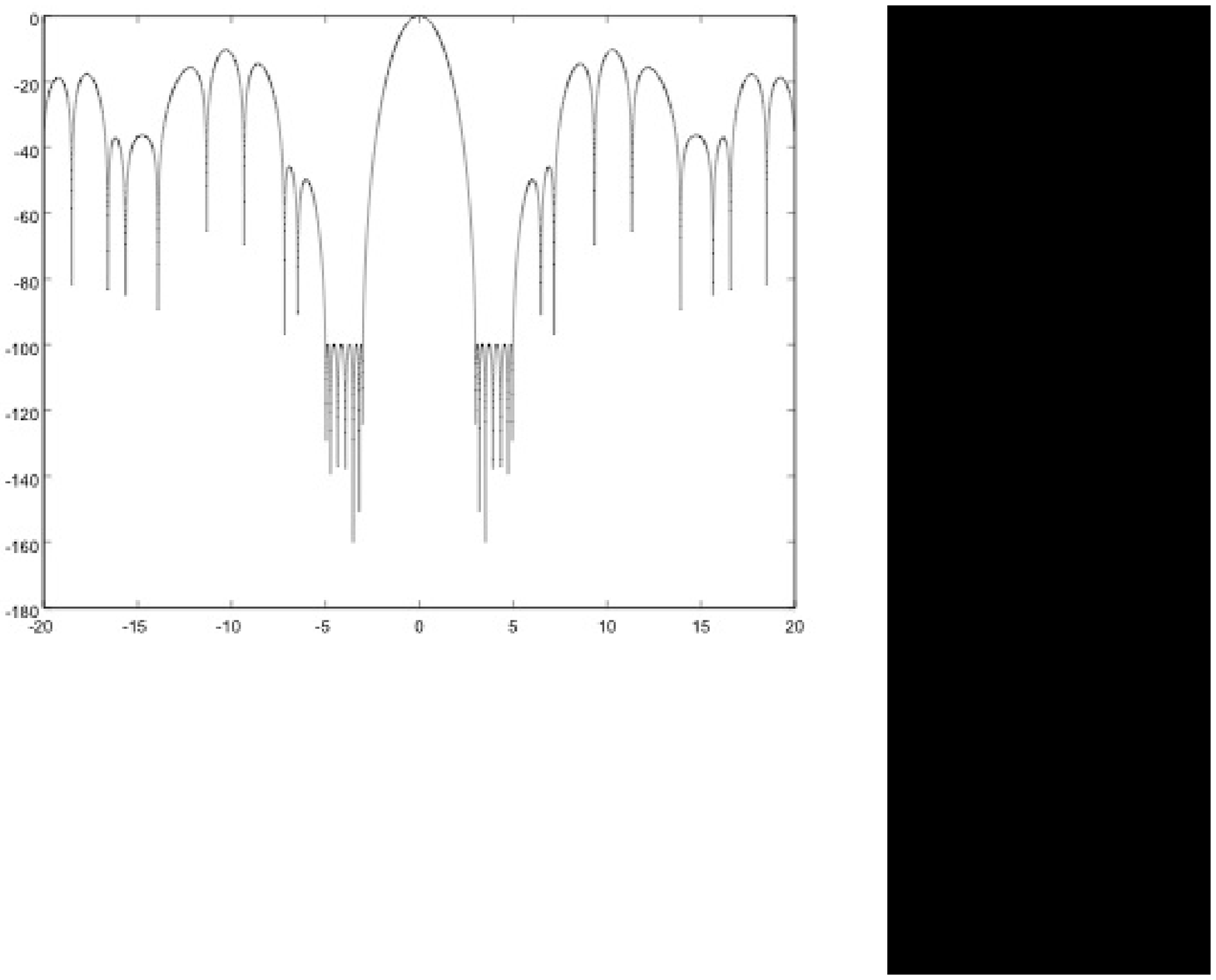}
\end{center}
\caption{{\em Left} The Point Spread Function of a barcode mask designed for high contrast in the range 3 to 5 $\lambda/D$.  {\em Right} The cross-section of the PSF.  This mask has an Airy throughput of 33.8 \%. }
\label{fig:3to5barcode}
\end{figure}

In order to investigate as many planets as possible (particularly at larger distances), there is a strong desire to achieve as small an inner working angle as possible.  One benefit of small inner working angle is a reduction in the size of the telescope required to image the full habitable zone of these stars.  One way to achieve smaller inner working angle is to optimize a narrower discovery region (that is, bring in the outer working angle).  \reffig{fig:3to5barcode} shows a cross section  of the PSF of such a design optimized for high contrast in the range 3 to 5 $\lambda/D$ with an Airy throughput of 33.8\%.  It also possible to design a mask for an inner working angle as small as $2\mbox{ } \lambda/D$, but the width becomes prohibitively small (2 to 2.2 $\lambda/D$).

Perhaps a more natural, but more complicated, optimization is to maximize the total throughput rather than the pseudo-area.  This results in a nonlinear optimization that is much more difficult to solve.  Nevertheless, we examined some solutions and found the same result as the pseudo-area problem.  We also note the importance of considering the Airy throughput for these finite band masks rather than the total throughput as the light gathered from a planet with this PSF would correspond only  to the inner core.  It is thus more appropriate to ignore the light scattered outside the outer working distance.

\begin{figure}[t]
\begin{center}
\plottwo{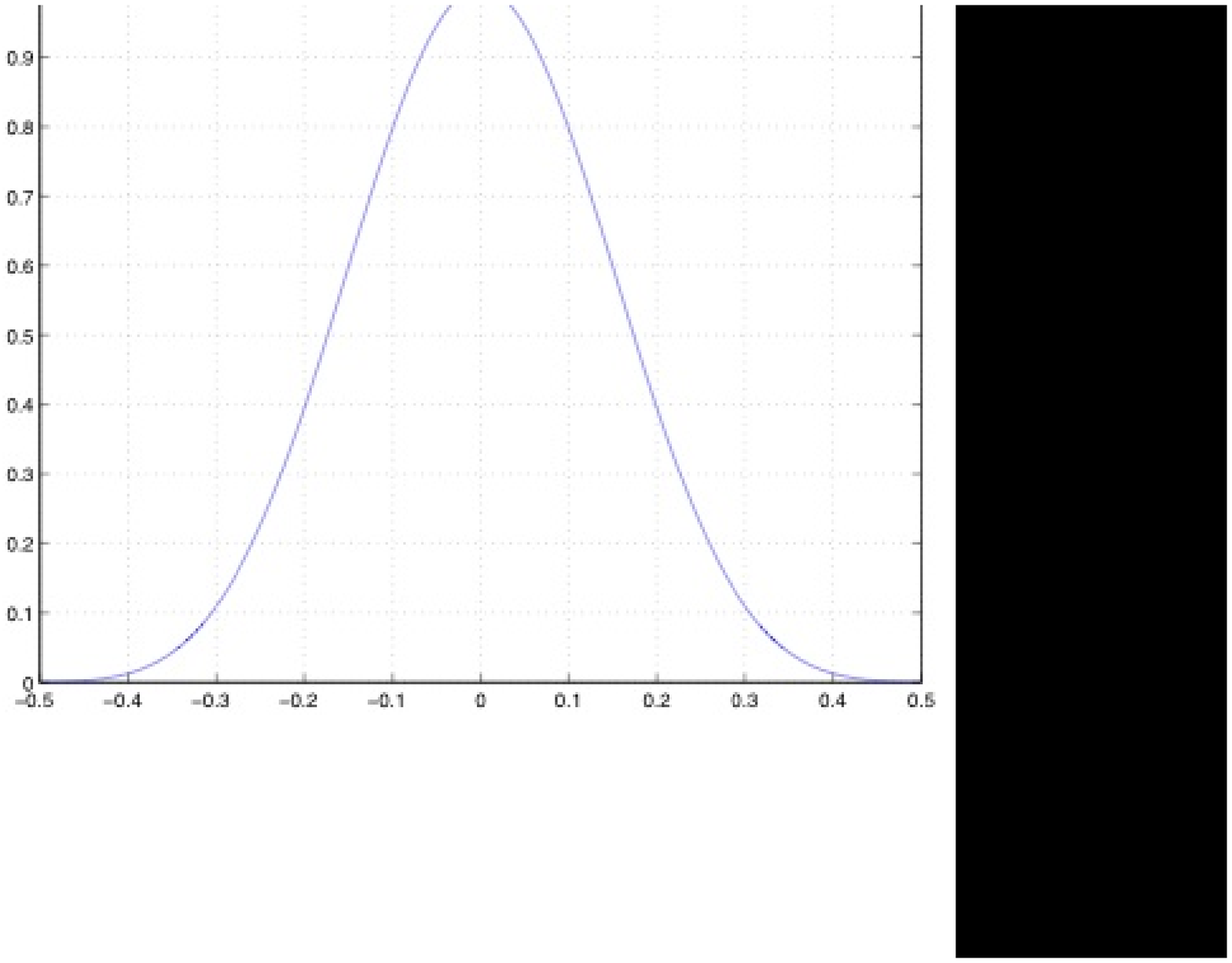}{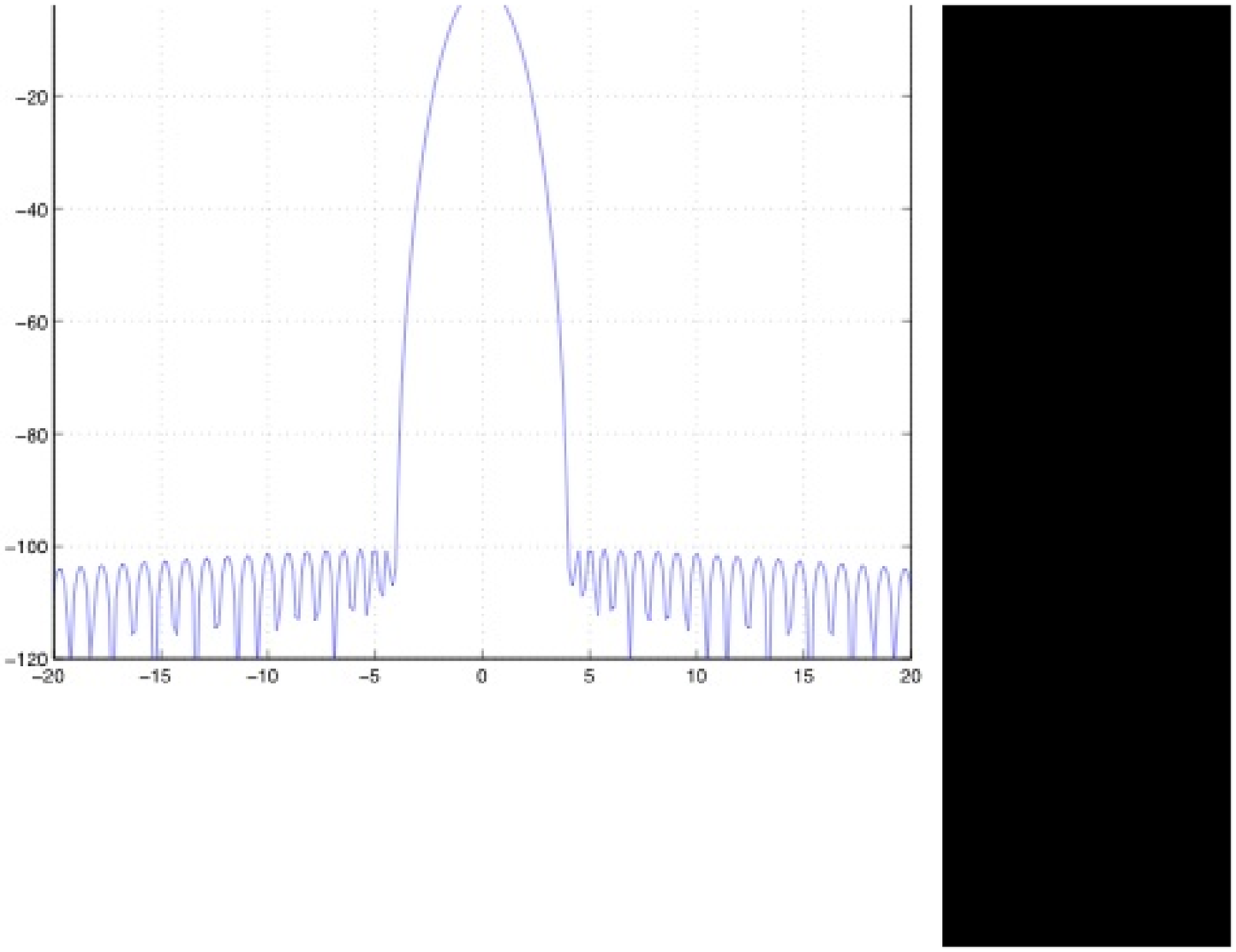}
\end{center}
\caption{{\em Left} A smooth one-dimensional optimal apodized function.
{\em Right} The corresponding cross section of the PSF showing an inner working distance of
$4 \lambda/a$ and contrast better than $10^{-10}$.}
\label{fig:OneDSmooth}
\end{figure}

Finally, it is possible to numerically obtain a smooth apodization similar to the one in \reffig{fig:OneDProlate} by modifying the original optimization problem.  To accomplish this, we add seemingly artificial smoothness constraints.  Motivated by the fact that optimal apodizations look qualitatively like a Gaussian function, we impose smoothness constraints that correspond to the following conditions:
\begin{eqnarray}
\log(A)'& \le & 0  \label{eq:logpcon} \\
\log(A)'' & \le & 0  \label{eq:logppcon}
\end{eqnarray}
The resulting max-pseudo-area optimization problem is:
\begin{equation} \label{120}
    \begin{array}{ll}
        \mbox{maximize } & \int_0^{1/2} A(x) dx \\
	\mbox{subject to } &
	    \setlength{\arraycolsep}{0.1em}
	    \begin{array}[t]{rcll}
	    \displaystyle
	        -10^{-5} E(0,\zeta) \le & E(0,\zeta) & \le 10^{-5} E(0,\zeta), &
	            \qquad \xinaught \le \xi \le \xione , \\
		  0 \le & A(x)   & \le  1, & \qquad 0 \le x \le 1/2, \\
		        & A'(x)  & \le  0, & \qquad 0 \le x \le 1/2, \\
		        & A(x) A''(x)& \le A'(x)^2, & \qquad 0 \le x \le 1/2. \\
	    \end{array}
    \end{array}
\end{equation}
\reffig{fig:OneDSmooth} shows the resulting smooth apodization function and the corresponding PSF (i.e., the Fourier Transform).  Not surprisingly, it has a strong similarity to the prolate spheroidal wavefunction.

\subsection{Phase Shifts in Apodized Masks}
\label{sec:phaseshifts}

The central drawback of apodized pupils, and one of the significant factors that led us to consider  binary masks, is the difficulty manufacturing them.   It is extremely hard to create a smoothly varying mask to the necessary accuracy for $10^{-10}$ contrast.  In addition, there is no known process for creating apodized masks that doesn't also introduce a phase shift as a function of the transmission.   This phase shift can dramatically reduce the performance of the apodized pupil.  To better understand the issue and in an effort to ameliorate the problem, we examined optimizing in the presence of such a phase shift.

Incorporating the effect of phase shift,
the correct expression for the electric field is
\[
    E(\xi) =  
    \int_{-1/2}^{1/2} e^{-2 \pi i x \xi + i k \log(A(x))} A(x) dx
\]
where $k$ is a constant giving the rate that lag appears as attenuation is
increased.
Note that, unlike our earlier expressions, 
this complex expression does not reduce to a real-valued one.
\begin{figure}[t]
\begin{center}
\includegraphics[width=3.0in, angle=90, origin=c]{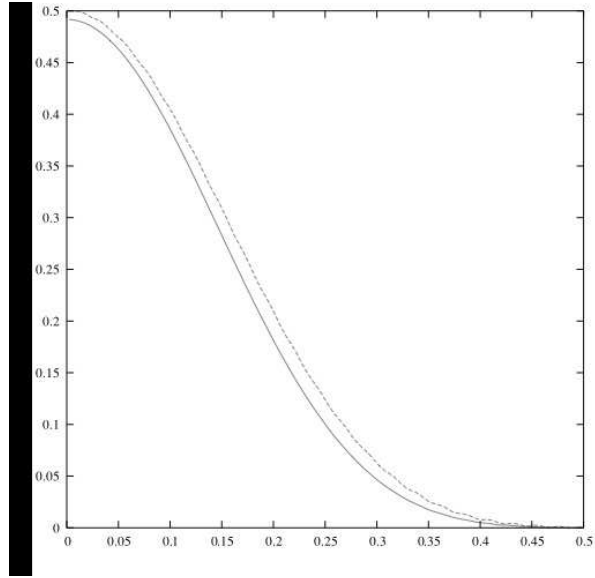} 
\end{center}
\caption{Optimal Apodization with and without Phase Shift}
\label{PhaseShiftApodizations}
\end{figure}
In \reffig{PhaseShiftApodizations},
the upper apodization was computed with $k=0$ (i.e., no phase shift) and
using $\xinaught = 4$ and $\xione = 40$.
The lower apodization corresponds to a nonzero phase shift.  It was computed
using a larger inner working angle of $\xinaught = 4.5$.  
The problem seems to be infeasible with a tighter inner working angle.

\begin{figure}[t]
\begin{center}
\includegraphics[width=3.0in, angle=90, origin=c]{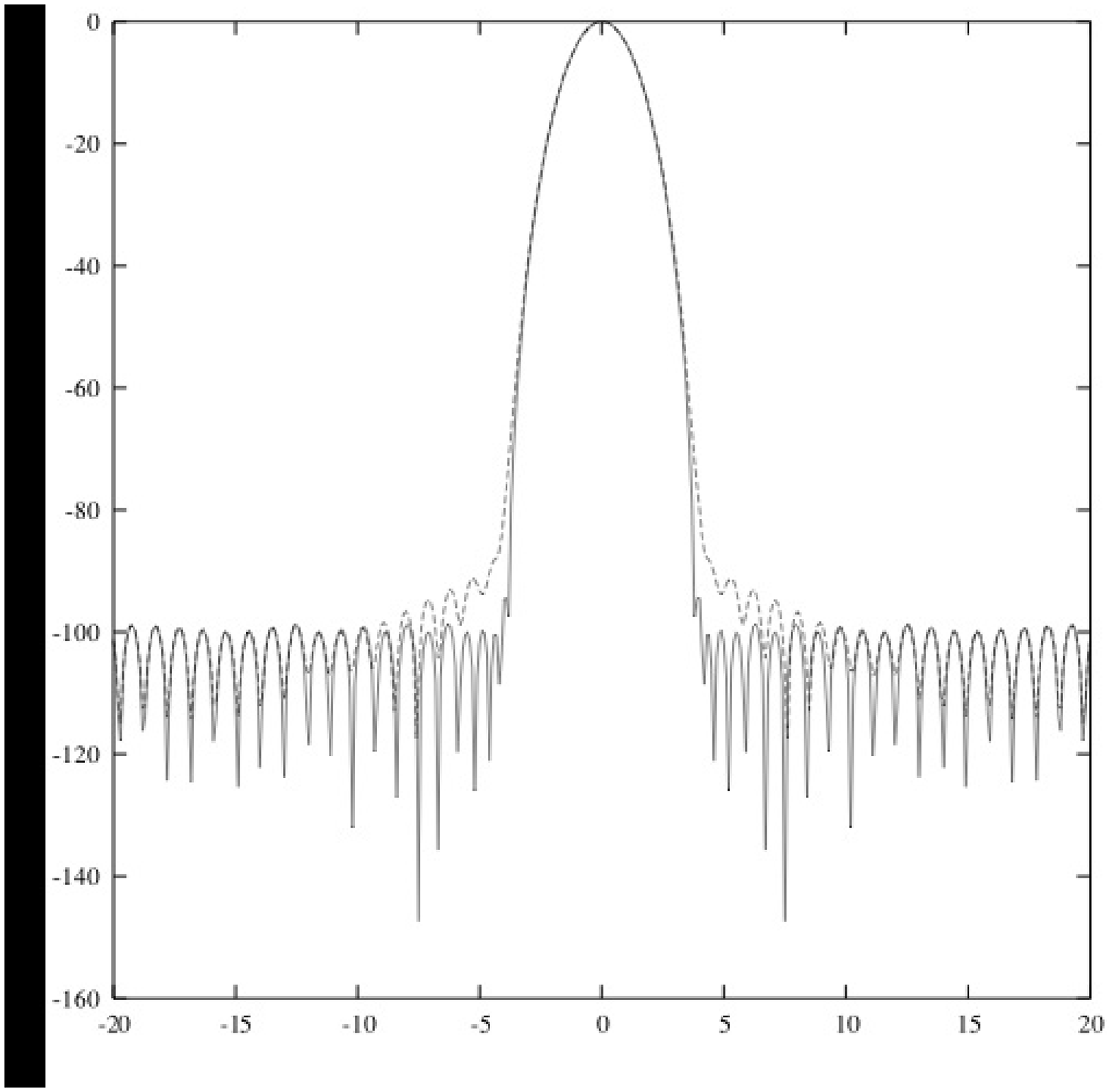} 
\end{center}
\caption{PSF for Optimal Apodization without Phase Shift.
Here is shown the PSF for the apodization computed without accounting for
phase shift.  
The upper curve is the PSF one gets with the phase shift term using $k=0.1$.
}
\label{nophaseshift}
\end{figure}

If the apodization is computed assuming no phase shift but in reality there is
a phase shift, then one expects the PSF to degrade because of the shift.
\reffig{nophaseshift} shows the amount of degradation one will see with $k=0.1$.

\begin{figure}[t]
\begin{center}
\includegraphics[width=3.0in, angle=90,origin=c]{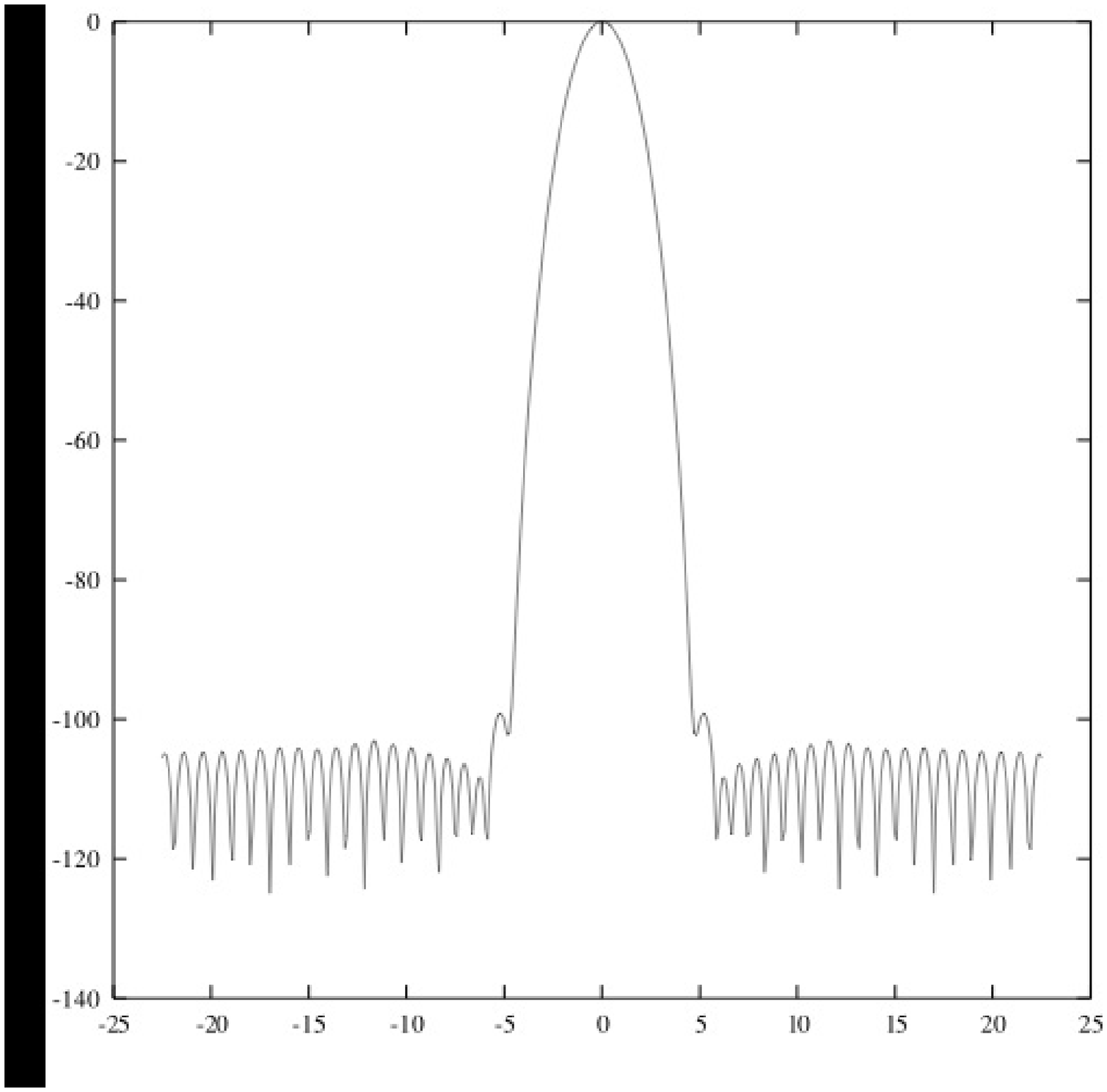} 
\end{center}
\caption{PSF for Optimal Apodization with Phase Shift.
Here is shown the PSF for the apodization computed with accounting for
phase shift.  Inner working angle is $4.5 \lambda/D$.
}
\label{withphaseshift}
\end{figure}

Finally, \reffig{withphaseshift} shows the psf corresponding to the apodization computed
assuming a phase shift of $k=0.1$.  Note that the iwa is now $4.5$.

\section{Asymmetric Shaped Pupils}

Because of the problems associated with manufacturing and the limitations due to phase shifts described above, apodized pupils present unique and difficult challenges as a mechanism for modifying the point spread function to achieve high contrast.  We favor, instead, shaped pupils, where manufacturing involves only cutting openings in the material at accuracies easily achieved with current technology.  We have already, in fact, presented one class of shaped pupils as the optimal solution to the high-contrast imaging problem---the barcode mask of Section~\ref{sec:one_d} (see also \cite{VSK03} and \cite{VSK02} for examples of circularly symmetric shaped pupil masks).  In the following subsections we present a collection of other asymmetric shaped pupil masks derivable from a variety of optimization problems. 

\subsection{Single Pupil}
David Spergel was the first to realize that shaped pupils could be used to achieve essentially arbitrarily deep nulls close to the central star.  His first suggestion (\cite{ref:Spergel}) was to use a gaussian-shaped mask.  Shortly thereafter, Jeremy Kasdin realized that Slepian's prolate-spheroidal wave function (\cite{ref:Slepian1}) could be used here too as the optimal shape for achieving high contrast along one axis of the image plane.  This is seen by returning to the expressions for apodizations in a square or rectangular aperture (Eqs.~\ref{eq:squareapod} and \ref{eq:onedapod}).  If in place of a smooth apodization we assume instead a single pupil opening of width $w(x)$ relative to the $y=0$ axis we have for the electric field on the $\zeta=0$ axis in the image plane:
\begin{equation}
E(\xi,0)=\int_{-1/2}^{1/2}{w(x)e^{-i2\pi\xi x}dx}
\end{equation}

\begin{figure}[ht]
\begin{center}
\includegraphics[width=2in]{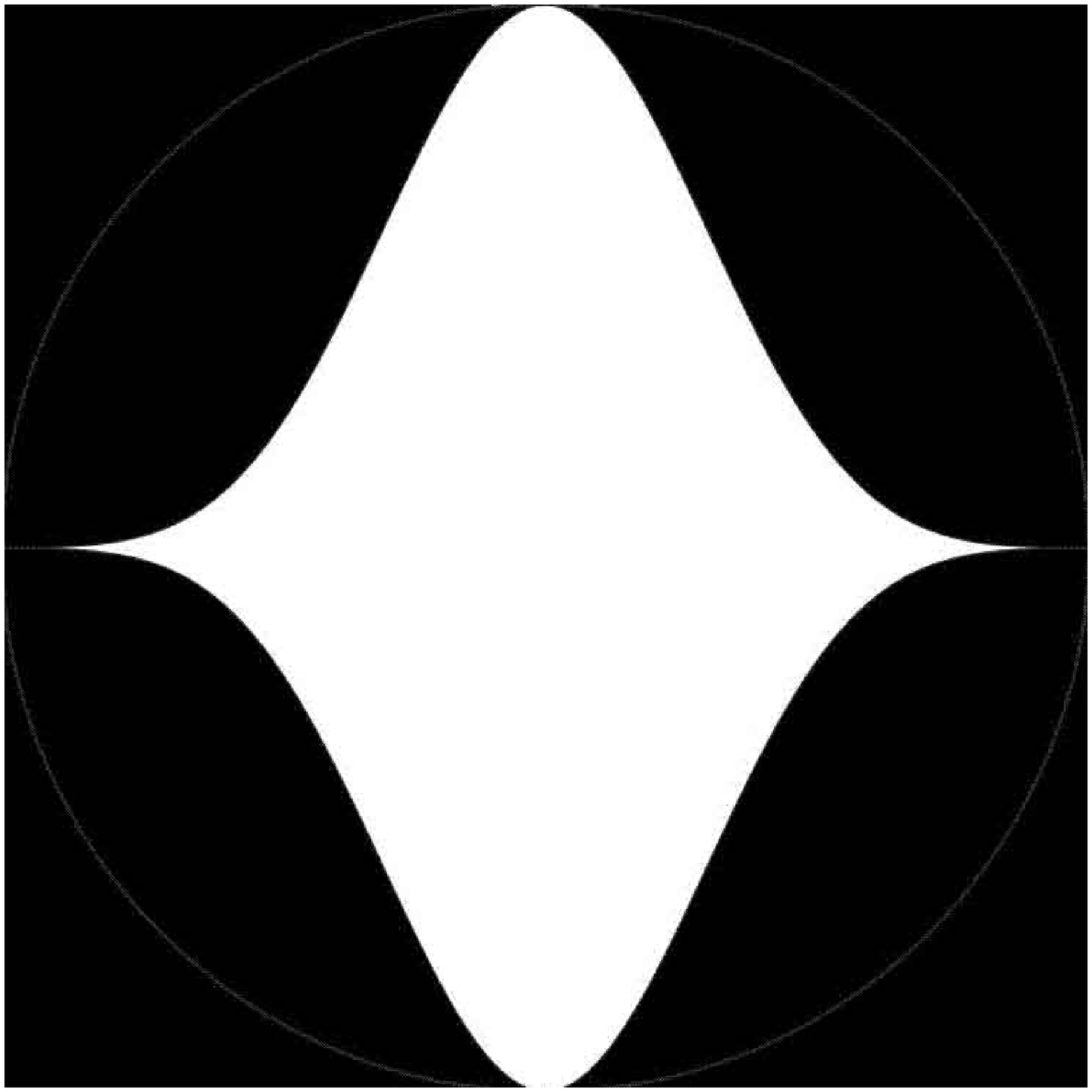} 
\includegraphics[width=2in]{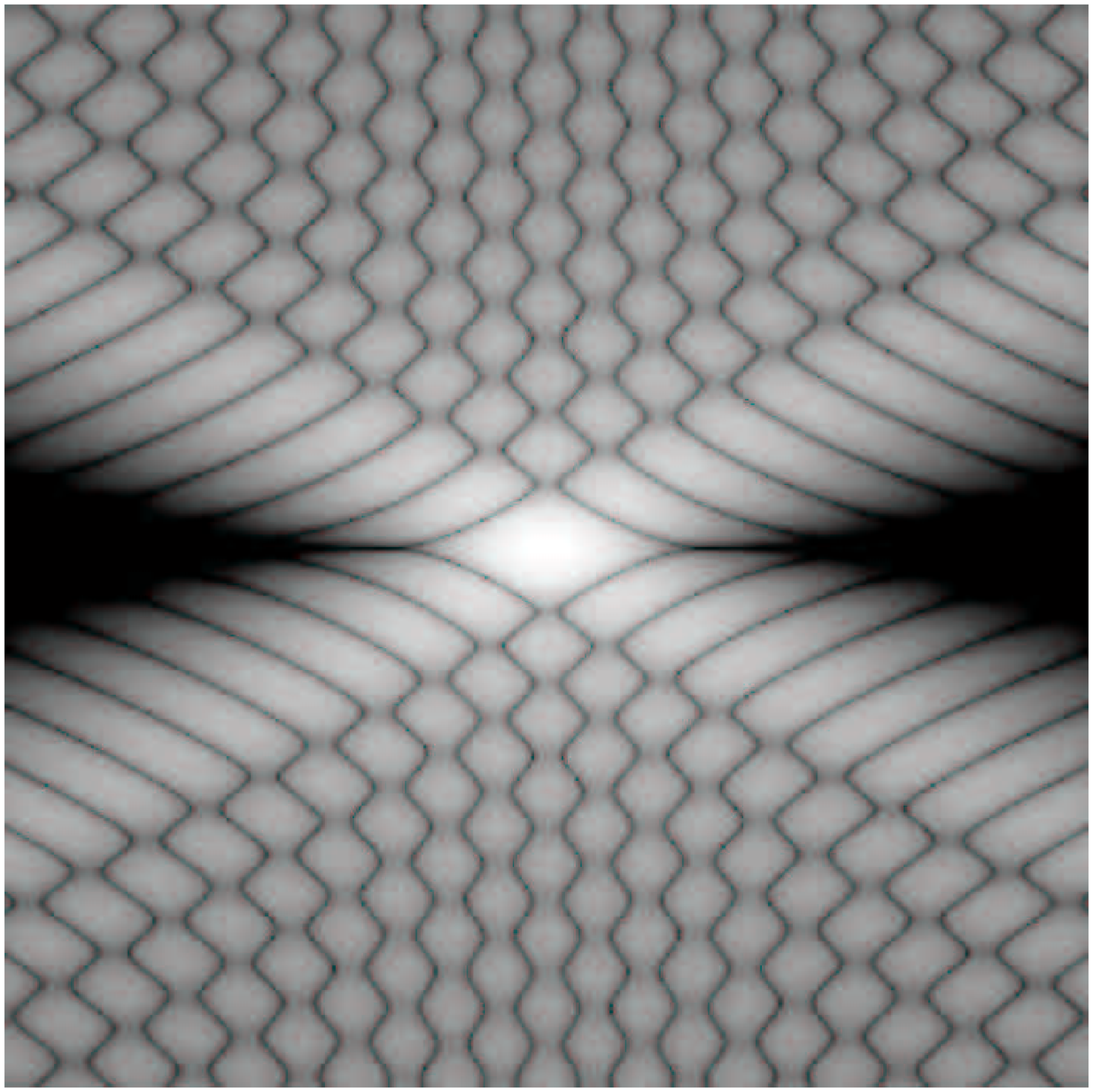}
\includegraphics[width=2.1in, angle=90, origin=c]{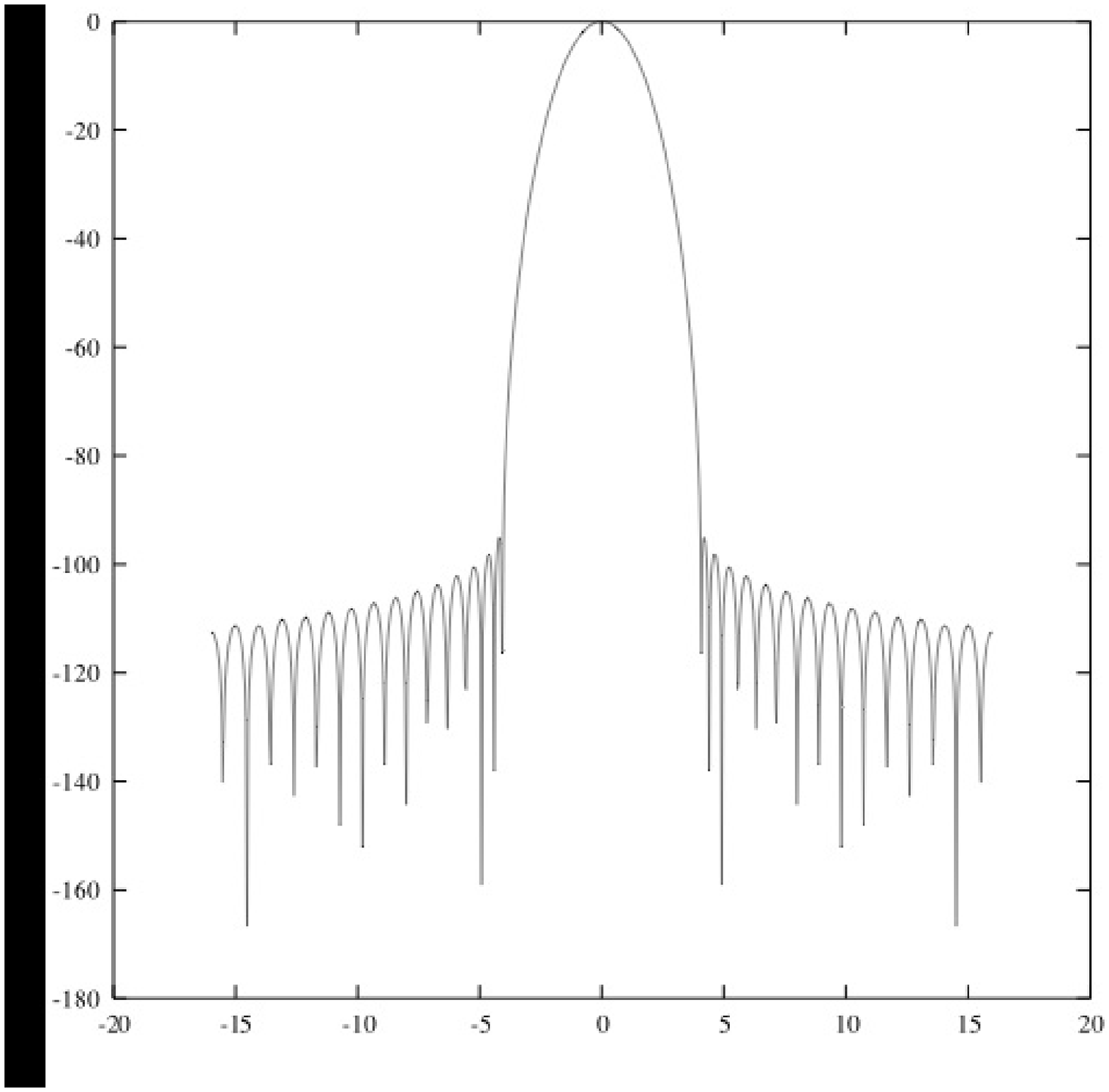}
\end{center}
\caption{{\em Top.} The Spergel-Kasdin prolate-spheroidal mask.  
{\em Bottom.} The associated
$2$-D point spread function and its $x$-axis slice shown in decibels
($10^{-10} = -100\mbox{dB}$).}
\label{fig1}
\end{figure}

In other words, the field on the axis of the image plane is just the one-dimensional Fourier Transform of the pupil width.  Thus, an optimization problem for that width is identical to the Slepian problem we described earlier for the one-dimensional apodization.  A single pupil with the Slepian shape would provide the needed contrast.  Such a pupil is shown in \reffig{fig1}.  It achieves a contrast of
$10^{-10}$ everywhere along the $x$-axis except within an Airy disk of radius $4 \lambda/D$.  This was our first proposed shaped pupil for TPF.

It is clear from \reffig{fig1} that the disadvantage of the Spergel-Kasdin mask is the extreme narrowness of the high-contrast region at close working angles from the central star.  Many rotations of the telescope would be necessary to image the entire discovery space, dramatically impacting integration time and mission life.  The solution to opening up this discovery space is to introduce multiple openings, what we call multi-pupil masks.  In fact, the barcode  mask introduced earlier is an example of a multi-pupil mask.    In what follows we describe several other approaches to accomplishing this end.

\subsection{Multi-Pupil Square Aperture}

The simplest possible multi-pupil mask involves just repeating the single pupil, one on top of another, in a square aperture until the desired PSF is achieved.  In this approach, rather than optimizing a design directly, we attempt to reproduce, as closely as possible, the PSF of the one-dimensional apodization described in Section \ref{sec:one_d} and shown in \reffig{fig:OneDProlate}.  To begin, we assume that each opening in the mask is identical and stacked vertically across the pupil.  The opening, $y(x)$, in the upper quadrant of a $2N$ pupil mask, can then be described by the set S given by:
\begin{eqnarray}
S & = & \{(x,y): -1/2 \le x \le 1/2, y(x) \in \mathrm{Y_u}\} \\
\mathrm{Y_u} & = & \bigcup_{n=0}^{N-1}{\left [\frac{D}{2}\frac{n}{N}+\frac{w(x)}{2}, \frac{D}{2}\frac{n+1}{N}-\frac{w(x)}{2} \right ] }
\end{eqnarray}
while for the lower pupils, $y(x) \in \mathrm{Y_l}$,
\begin{equation}
\mathrm{Y_l}  =  \bigcup_{n=0}^{N-1}{\left [-\frac{D}{2}\frac{n+1}{N}+\frac{w(x)}{2}, -\frac{D}{2}\frac{n}{N}-\frac{w(x)}{2} \right ] }
\end{equation}
where $w(x)$ denotes the half width of the pupil and $D$ is the horizontal width of the aperture.  For what follows, all lengths  are normalized by $D$ to simplify notation.  Note that the shape of each pupil is the same and determined by $w(x)$.  Again, the goal is to determine choices for this function that will yield an image-plane PSF matching the PSF corresponding to a given apodization (such as that in \reffig{fig:OneDProlate}).  The electric field for such a multi-opening mask can by found from \refeqn{eq:onedapod}:
\begin{equation}
E(\xi,\zeta)=\frac{2\sin(\pi\zeta)}{\pi\zeta}\int_0^{1/2}\sum_{n=0}^{N-1}{{\left [ \sin 2\pi\zeta\left(\frac{n+1}{2N}-\frac{w(x)}{2}\right) - \sin2\pi\zeta\left(\frac{n}{2N}+\frac{w(x)}{2}\right)\right]\frac{\cos(2\pi \xi x)}{\sin(\pi\zeta)} dx }}
\end{equation}
which can be simplified by completing the finite sum:
\begin{equation}
E(\xi,\zeta)=\frac{2\sin(\pi\zeta)}{\pi\zeta}\int_0^{1/2}{\left ( \cos(\pi \zeta w(x)) - \frac{[1+\cos(\pi \zeta/N)]}{\sin(\pi \zeta/N)}\sin(\pi \zeta w(x)) \right ) \cos(2\pi \xi x) dx}
\label{eq:Ew}
\end{equation}

This looks  like the field due to an arbitrary apodization $A(x)$ in \refeqn{eq:onedapod}, except with $A(x)$ replaced by the expression in parentheses.  If we expand this expression in a Taylor series about $\zeta=0$, we find that the leading term in the series is  $1-2Nw(x)$, independent of $\zeta$.  Thus, if we choose:
\begin{equation}
w(x) = \frac{1-A(x)}{2N}
\label{eq:w(x)}
\end{equation}
we can create an approximation to the continuous one-dimensional apodization response.  What remains is to show that the higher order terms, for large enough $N$, only produce interfering light outside some outer working angle $\zeta_{\mbox{\scriptsize owa}}$.  Replacing for $w(x)$ from \refeqn{eq:w(x)} into \refeqn{eq:Ew} results in:
\begin{equation}
E(\xi,\zeta)=\frac{2\sin(\pi\zeta)}{\pi\zeta}\int_0^{1/2}{\left (\frac{\sin\left (\frac{\pi \zeta (1+A(x))}{2N} \right ) - \sin \left (\frac{\pi \zeta (1-A(x))}{2N} \right )}{\sin(\pi \zeta /N)} \right ) \cos(2\pi \xi x) dx}
\label{eq:EA}
\end{equation}

We can now Taylor expand the term in parentheses in \refeqn{eq:EA} for small $(\pi \zeta/N)$ to find:
\begin{equation}
E(\xi,\zeta)=\frac{2\sin(\pi\zeta)}{\pi\zeta}\int_0^{1/2}{\left (A(x) + \mathcal{O}\left ( \frac{\pi \zeta}{N} \right )^2 \right ) \cos(2\pi \xi x) dx}
\end{equation}

\begin{figure}[h]
\plottwo{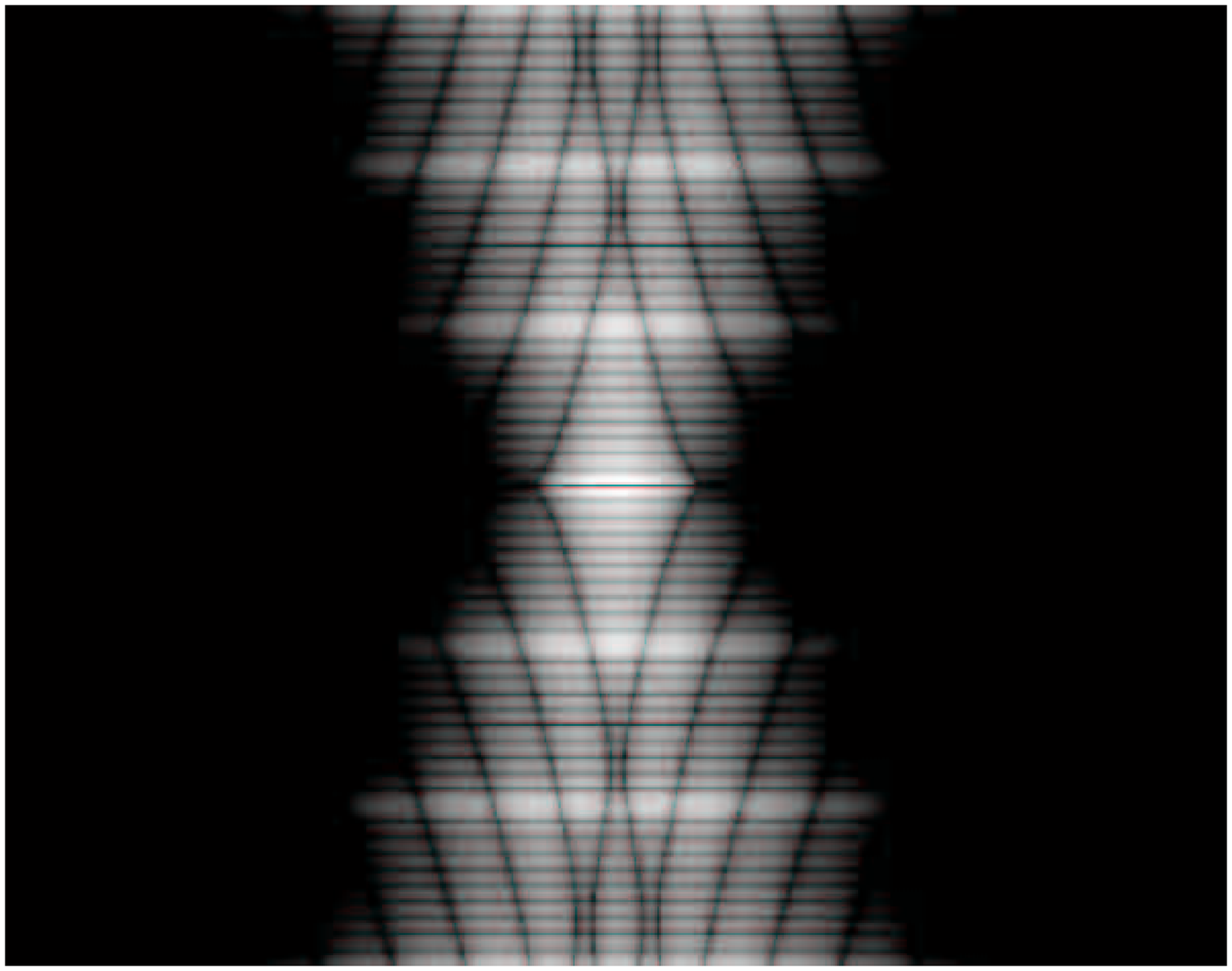}{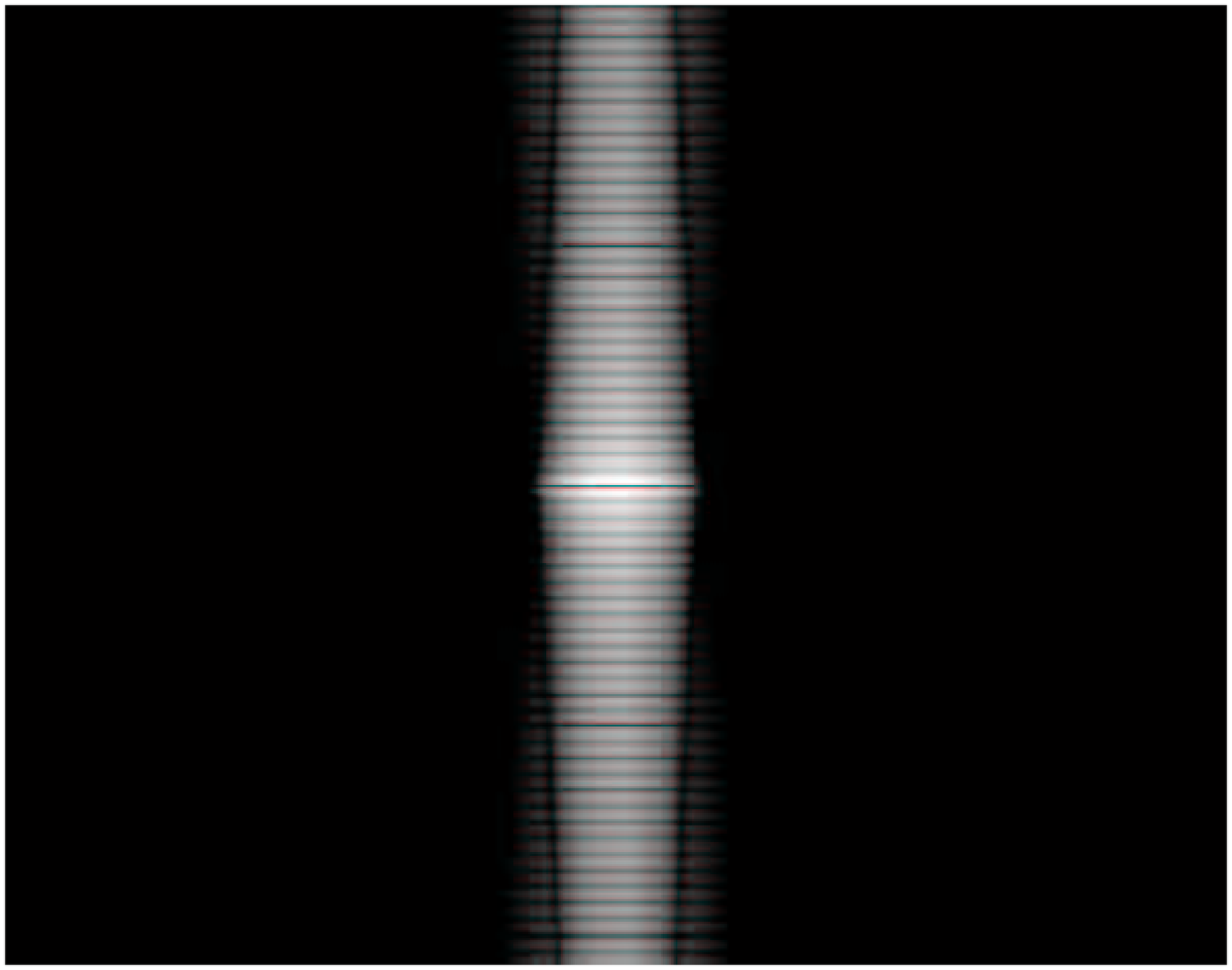}
\caption{PSFs for a multipupil square aperture associated with the apodization shown in \reffig{fig:OneDProlate}.  {\em Left} A 10 opening pupil.
{\em Right} A 100 opening pupil.}
\label{fig:multisame}
\end{figure}

Thus, for large enough N, the scattered light due to the finite number of openings can be made arbitrarily small within some range 
 $0\le\zeta\le\zeta_{\mbox{\scriptsize owa}}$.  \reffig{fig:multisame} shows the PSF resulting from such a multi-opening square aperture with 10 and 100 openings using the prolate spheroidal apodization shown in \reffig{fig:OneDProlate} for $A(x)$.

While it is certainly feasible, and perhaps not overly difficult, to manufacture a mask with over 100 openings, it may be desirable to use fewer pupils.  It is thus sensible to attempt to improve the resopnse with fewer openings by relaxing the requirement that each edge of the openings be the same, hoping  that these extra degrees of freedom can be used to improve the match to the apodized pupil, albeit at the cost of a  much more difficult computational problem.

In this case, the set $\mathrm{Y_u}$ for the upper half of the pupil is replaced by:
\begin{equation}
\mathrm{Y_u}  =  \bigcup_{n=0}^{N-1}{\left [\frac{1}{2}\frac{n}{N}+\frac{b_n(x)}{2}, \frac{1}{2}\frac{n+1}{N}-\frac{t_n(x)}{2} \right ] }
\end{equation}
where $b_n(x)$ is the bottom edge of opening $n$ and $t_n(x)$ is the top edge of opening $n$.  A similar set holds for the lower half of the pupil.  The resulting electric field in the image plane is given by:
\begin{equation}
E(\xi,\zeta)=\frac{2\sin(\pi\zeta)}{\pi\zeta}\int_0^{1/2}\sum_{n=0}^{N-1}{{\left [ \sin \pi\zeta\left(\frac{n+1}{N}-t_n(x)\right) - \sin\pi\zeta\left(\frac{n}{N}+b_n(x)\right)\right]\frac{\cos(2\pi \xi x)}{\sin(\pi\zeta)} dx }}
\end{equation}

\begin{figure}[h]
\begin{center}
\includegraphics[width=2 in]{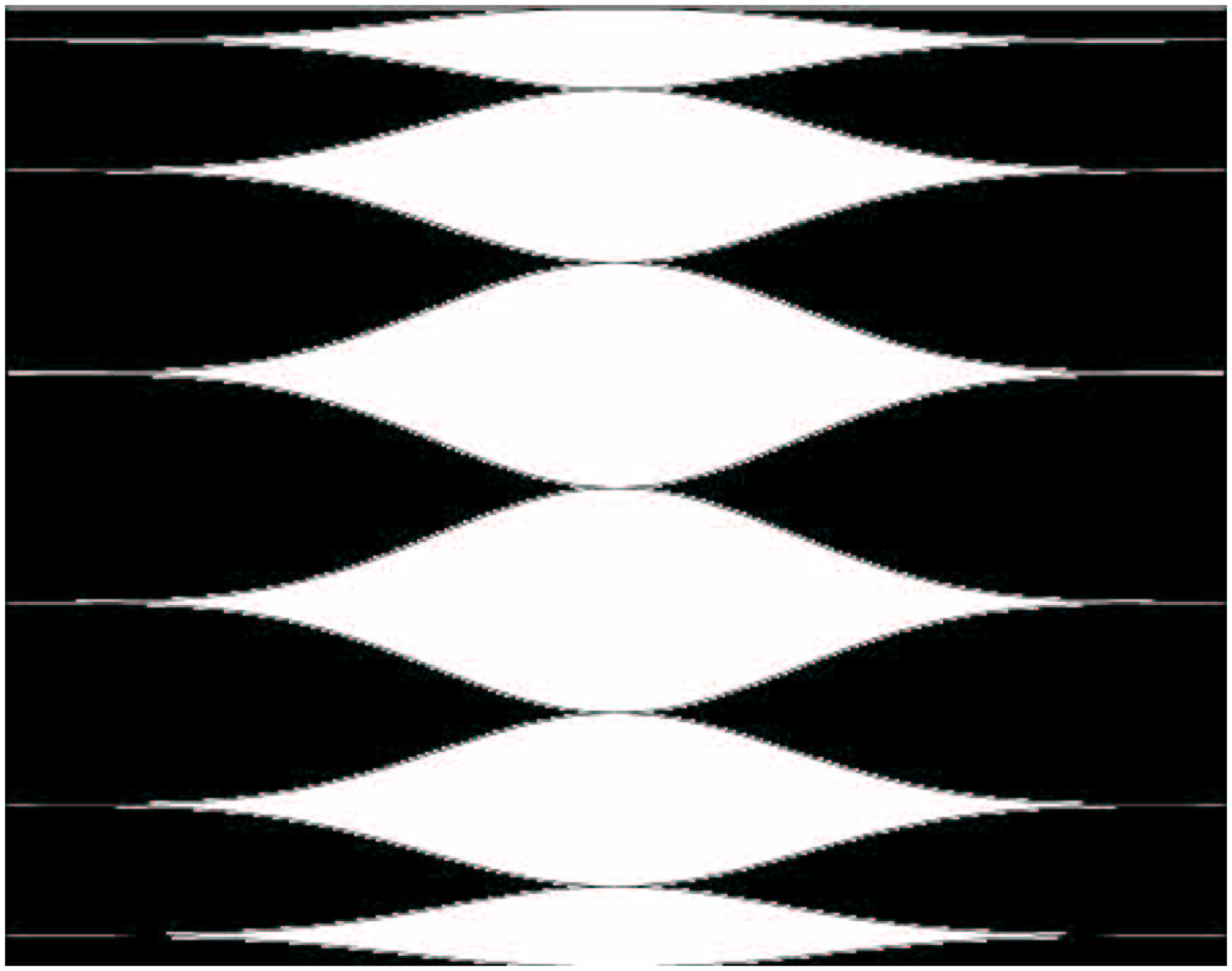}
\includegraphics[width=2 in]{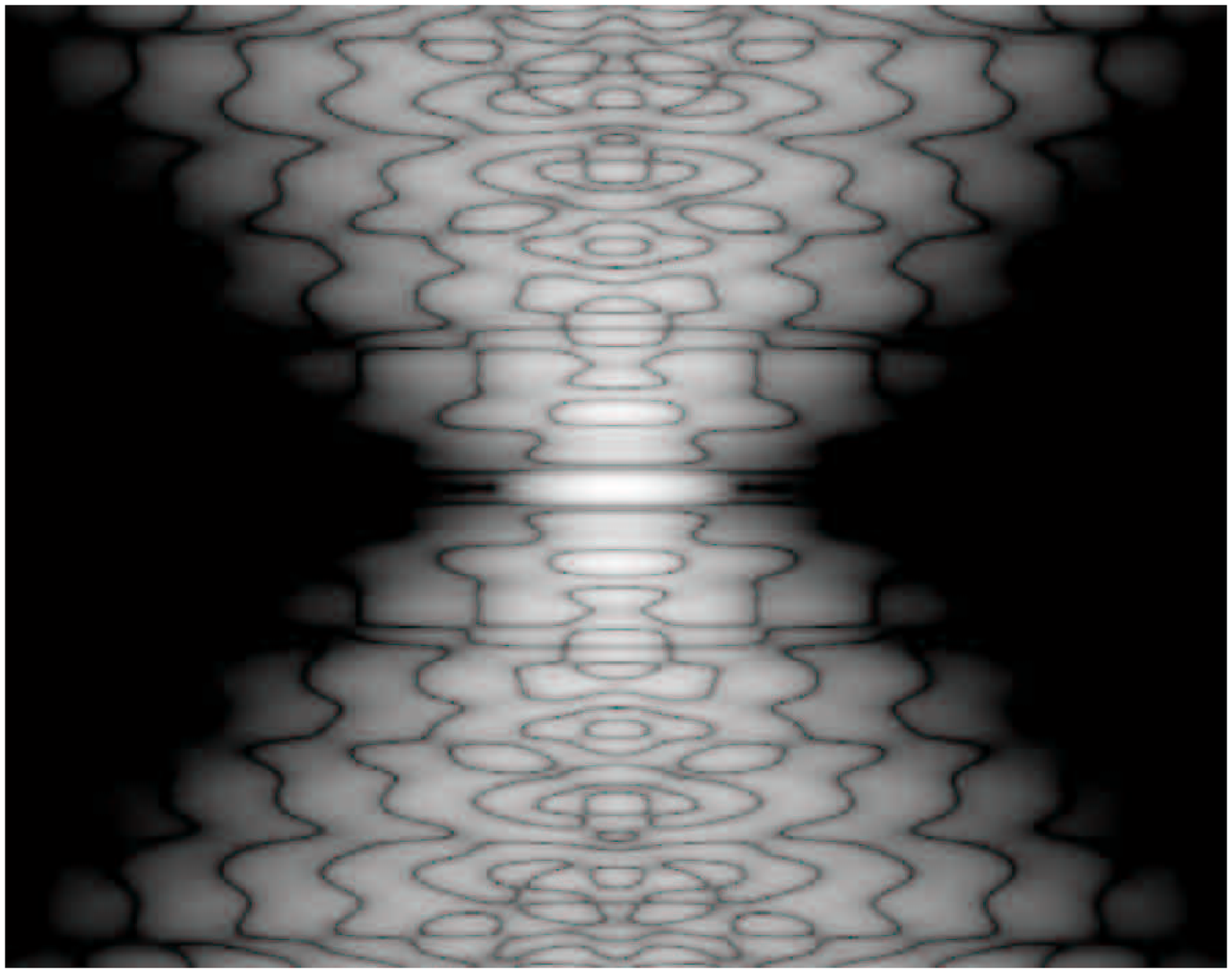}
\includegraphics[width=2 in]{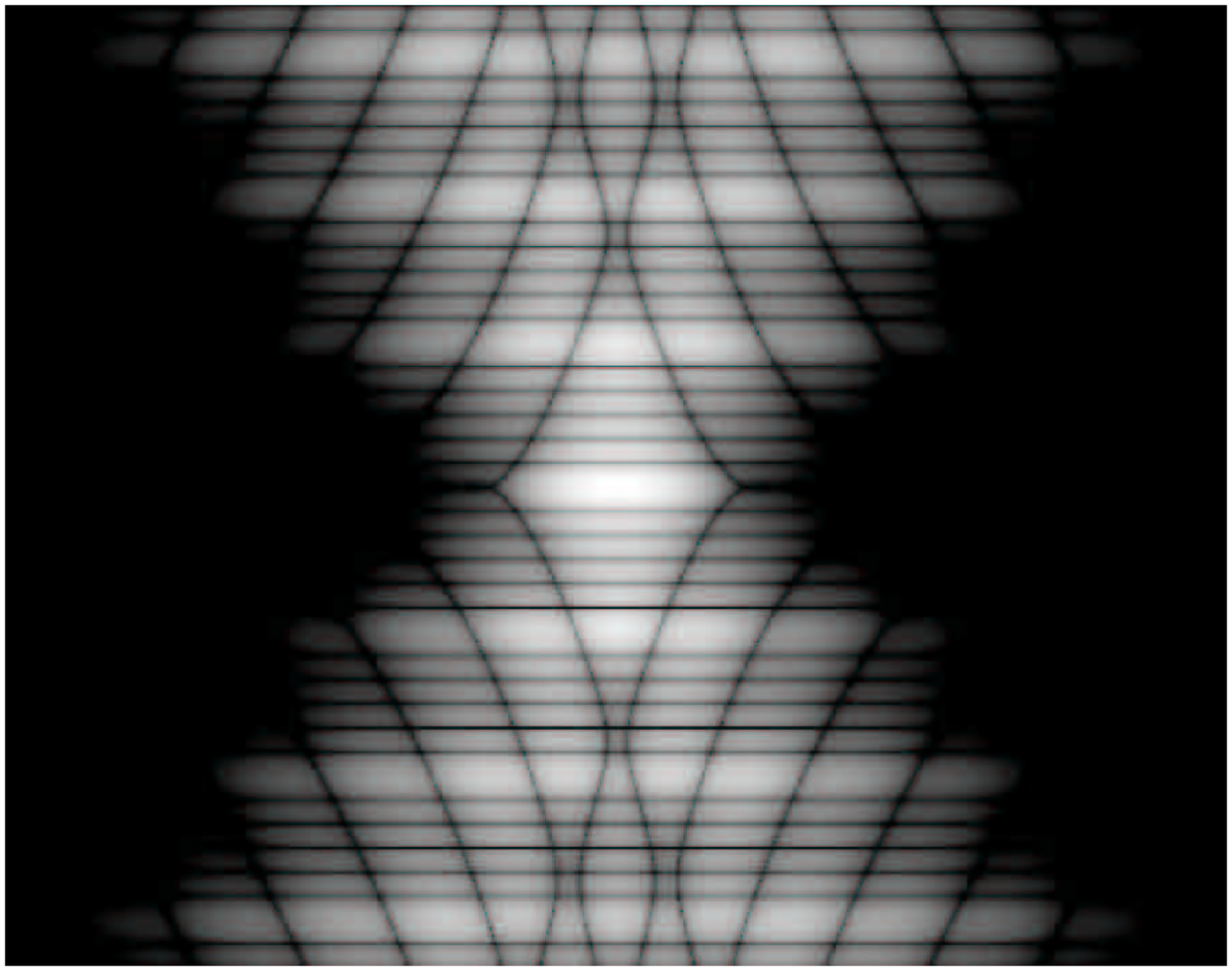}
\end{center}
\caption{{\em Left} An unequal six-opening pupil inscribed in a square aperture. 
{\em Center} The Corresponding PSF plotted on a logarithmic scale with black areas $10^{-10}$ below brightest. {\em Right} The PSF of a 6 equal opening multi-pupil aperture.  Spatial Frequencies shown from -20 to 20 $\lambda/D$.}
\label{fig:6pupil}
\end{figure}

Comparing again to \refeqn{eq:onedapod} we find that making the correspondence:
\begin{equation}
A(x) = \sum_{n=0}^{N-1}{\frac{\sin \pi\zeta\left(\frac{n+1}{N}-t_n(x)\right) - \sin\pi\zeta\left(\frac{n}{N}+b_n(x)\right)}{\sin\pi\zeta}}
\label{eq:match}
\end{equation}
results in a multi-pupil mask reproducing the one-dimensional apodization.  We again expand the right-hand side of \refeqn{eq:match} about $\zeta=0$ and match the lowest order term (independent of $\zeta$):
\[
A(x) = 1-\sum_{n=0}^{N-1}t_n-\sum_{n=0}^{N-1}b_n
\]
Unlike the previous multi-pupil mask, we can now use the extra degrees of freedom to make the coefficients of the terms of higher order in $\zeta$ vanish.  Each additional pupil pair allows us to eliminate two higher order terms.  The computational problem to find  the shapes of the openings becomes the solution to  N simultaneous polynomial equations.   While intuitively  one would expect such extra degrees of freedom to result in a better match  to the apodized solution with fewer pupils, the surprising result is that no significant improvement is made.  \reffig{fig:6pupil} shows an example of a 6 opening pupil designed with this method, its PSF, and the PSF of a uniform 6 opening pupil. 

One of the advantages of asymmetric masks is that they  can be made to have higher throughput by reducing the size of the high contast region (smaller opening angle).  While this may be unattractive for planet detection, as many rotations become necessary to cover the discovery region, it can have large advantages for characterizing a planet at a known location.  We summarize another category of assymetric masks ideally suited to this problem in the next section.

\subsection{Multi-Pupil Circular and Elliptical Apertures}

\begin{figure}[h]
    \begin{center} 
    \plottwo{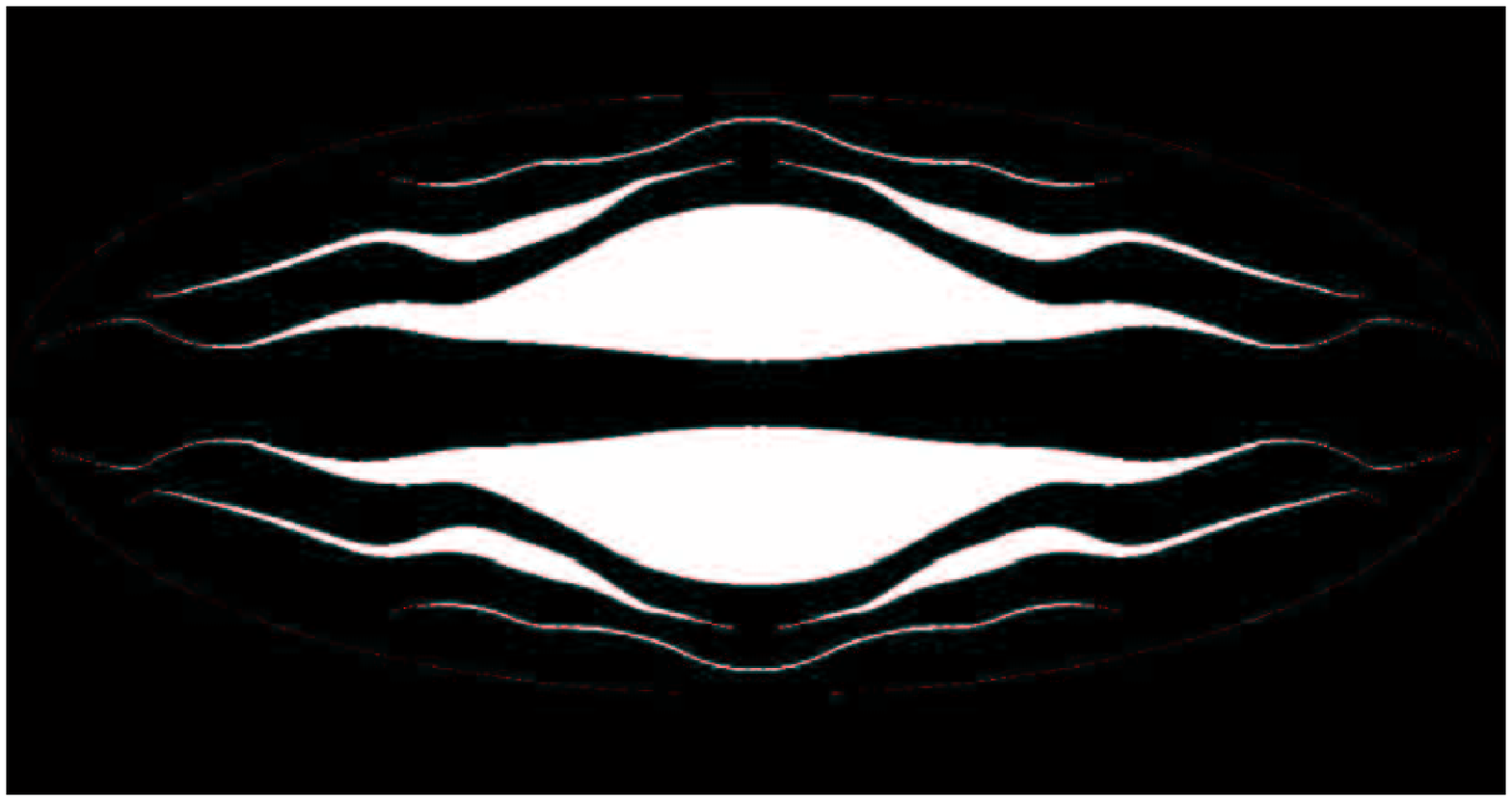}{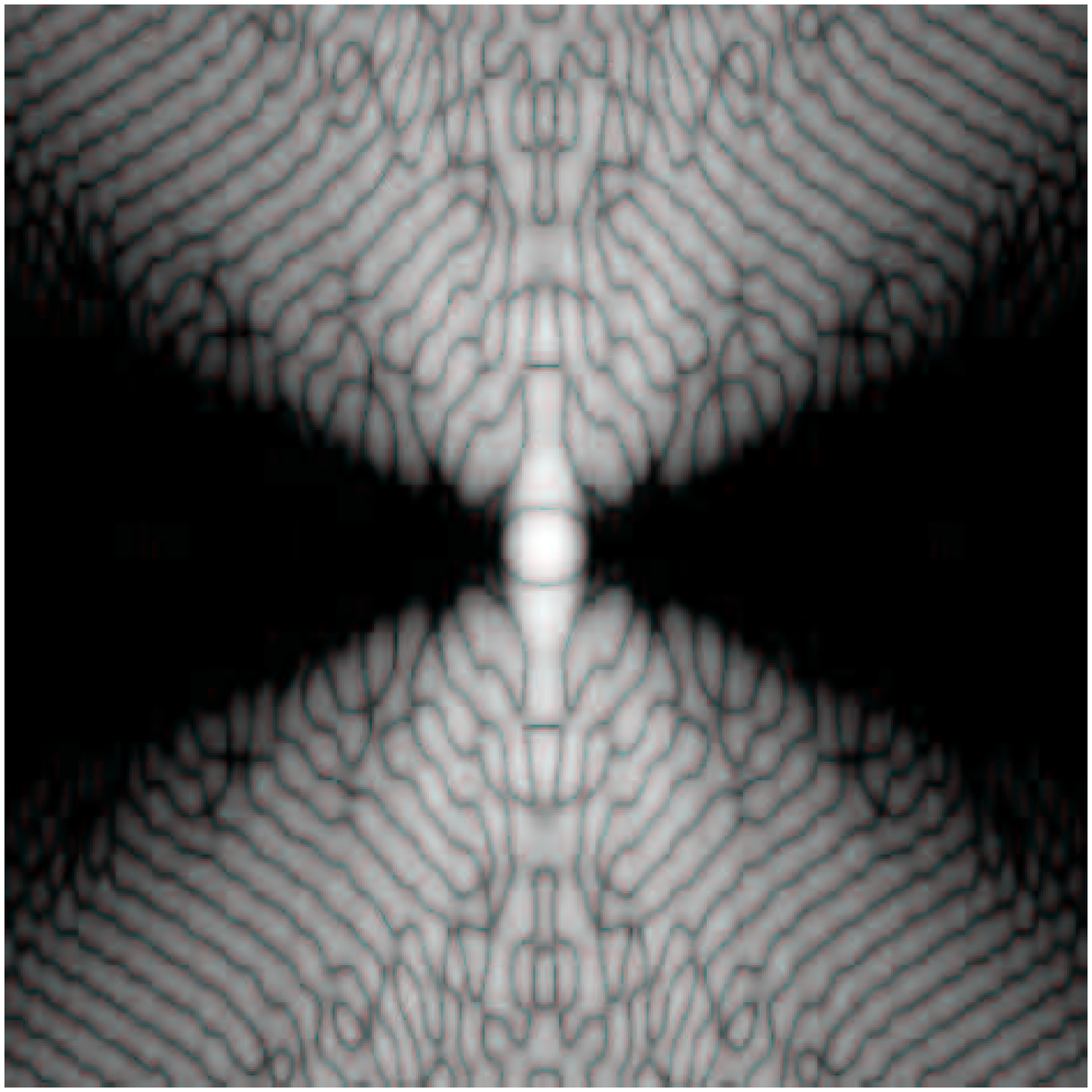}
    \end{center}
    \caption{{\em Left.} A multiopening pupil mask designed to open up the
    high-contrast region.  {\em Right.} The corresponding PSF.
    On the $x$-axis, contrast of $10^{-10}$ extends over $\rho \ge 4$.
    } 
    \label{fig:mult}
    \end{figure}

The multi-pupil asymmetric masks above are all embedded in square or rectangular apertures.  Thus, the central core of the PSF is also square, resulting in slight degradation of the inner working angle off axis.  Alternatively, multi-pupil assymetric masks can be designed that are embedded in circular and elliptical apertures by directly optimizing the edges of each opening relative to a performance function similar to those described in Section \ref{sec:optimal_apod}.   Many such pupils are described in detail in \cite{ref:KVSL}.  \reffig{fig:mult} shows an example of a multi-openning pupil mask that was designed for an elliptical aperture.  In an elliptical aperture, the asymmetry is used to leverage the long axis for improved iwa along the corresponding axis in the image plane.  
%
    
\section{Tolerances}
    
    A critical question  for any apodized or shaped pupil coronagraph is how well the mask needs to be made.  In this section we summarize the tolerancing requirements on an apodized pupil coronagraph and a barcode mask, insisting that any error  produce scattered light below the $10^{-10}$ contrast level.
    
    \subsection{Apodized masks}
    
    The electric field in the image plane for an apodized entrance pupil with some arbitrary error in amplitude and phase is given by a modification of \refeqn{eq:squareapod}:
\begin{equation}
E(\xi,\zeta)=\int{\int_{S}{e^{\gamma(x,y)}A(x,y)e^{-2\pi i (x\xi+y\zeta)}dxdy}}
\label{eq:apoderror}
\end{equation}
where $\gamma(x,y)=\alpha(x,y)+i \phi(x,y)$ is an arbitrary complex function of position in the pupil representing the phase and amplitude transmission error of the mask.  For simplicity, we consider here only the amplitude error in the apodization; we briefly discussed phase errors in Section~\ref{sec:phaseshifts}.  While the small amplitude error is arbitrary, because of the finite extent of the pupil we can model the error as a periodic function with fundamental period on $[-1/2, 1/2]$ (absorbing the top hat function into the apodization $A(x,y)$).  This allows us to decompose the error into a Fourier Series:
\begin{equation}
e^{\alpha(x,y)} = 1+\epsilon \sum_{n=1}^{\infty}{a_n \cos(\vec{k}_n \cdot \vec{x} + \varphi_n)}
\label{eq:ampfourier}
\end{equation}
where $\vec{x} \equiv (x,y)$ is the position vector, $\vec{k}_n = (k_x,k_y)$ is the wave vector of the harmonic term (with $|\vec{k}_n|$ being the spatial frequency), $A_n$ and $\varphi_n$ are the amplitude and phase of each harmonic component, and $\epsilon$ is small.  This expansion allows us to write the electric field in the image plane as the nominal, designed for point spread function plus a small  perturbation, $E(\xi,\zeta) = E_0(\xi,\zeta) + \Delta E(\xi,\zeta)$.  Since $\Delta E$ is small, the intensity in the image plane is approximately given by:
\begin{equation}
|E(\xi,\zeta)|^2 \approx |E_0(\xi,\zeta)|^2+ 2|E_0(\xi,\zeta)||\Delta E(\xi,\zeta)| 
\label{eq:Eerror}
\end{equation}

Since the electric field in \refeqn{eq:Eerror} is linear in the small change in intensity, superposition holds and it suffices to consider only a single harmonic component of the apodization error in \refeqn{eq:ampfourier}.  Thus, the field error is given by:
\begin{equation}
\Delta E = \epsilon \int{\int_{S}{A_n \cos(\vec{k}_n \cdot \vec{x} + \varphi_n)A(x,y)e^{-2\pi i (x\xi+y\zeta)}dxdy}}
 \end{equation}
 By the Fourier Shift theorem, this is just a copy of the point spread function to the location $(k_x, k_y)$ in the image plane, commonly referred to as speckle:
 \[ \Delta E = \epsilon E_0(\xi-k_x, \zeta-k_y) \]
 The intensity error in the high  contrast region, due to errors in manufacturing the apodization,  is thus given by:
 \begin{equation}
 \Delta_A \cong 2\epsilon |E_0(\xi,\zeta)| |E_0(\xi-k_x,\zeta-k_y)| \leq 10^{-10} |E_0(0,0)|^2
 \label{eq:delta_A}
 \end{equation}

 The requirement that $\Delta_A$ be less than $10^{-10}$ of the nominal PSF (below  the  contrast level) allows us to set a requirement on $\epsilon$, the magnitude of the apodization variation.  Since we are concerned only with the error in the high contrast region of the image plane, we can set $|E_0(\xi,\zeta)|/|E_0(0,0)|$ to $10^{-5}$ in \refeqn{eq:delta_A}, leaving:
 \begin{equation}
 \epsilon \leq \frac{5 \times 10^{-6}|E_0(0,0)|}{|E_0(\xi-k_x,\zeta-k_y)|}
 \label{eq:ap_epsilon}
 \end{equation}

 At the worst case error location of the speckle center, the ratio in \refeqn{eq:ap_epsilon} is unity, leaving a requirement on the manufacturing error of $\epsilon \le 5 \times 10^{-6}$.  Achieving a throughput variation with this accuracy is a formidable challenge.  This, in part, motivates our search for shaped pupil solutions.
 
 \subsection{Barcode Masks}
 
 While it is demonstrably easier to manufacture a shaped pupil mask than an apodized one, it is still necessary to estimate the precision required.  In this section we calculate the performance sensitivity of the barcode mask to manufacturing errors. Here we assume that the edge locations of each slot along the x-axis, $r_i$, are in error by some function $\epsilon_i f_i(y)$ on the right half and $\epsilon'_i f'_i(y)$ on the left half of the mask, where again $\epsilon$ is small.  We further assume that each $\epsilon_i$ and $\epsilon'_i$ are independent and identically distributed (i.i.d.) random variables with $\mathscr{E}\{\epsilon_i\} = 0$ and $\mathscr{E}\{\epsilon_i \epsilon_j \} = \sigma^2 \delta_{ij}$.  
 
 The error in the electric field is then given by:
 \begin{equation}
 \Delta E(\xi,\zeta) = \sum_{i=1}^N{\int_{-1/2}^{1/2}{e^{-2\pi i \zeta y} \left [ (\epsilon_i f_i(y) - \epsilon'_i f'_i(y) )\sin(2\pi r_i \xi) + i(\epsilon_i f_i(y) + \epsilon'_i f'_i(y))\cos(2\pi r_i \xi) \right ] dy}}
 \label{eq:deltaE_slot}
 \end{equation}
 Here, rather than use \refeqn{eq:Eerror} for  the image plane error we instead turn  to a stochastic analysis.  Since $\mathscr{E}\{\Delta E\}$ is zero, we compute the expected value of the intensity:
 \begin{equation}
 \mathscr{E}\{|E(\xi,\zeta)|^2\} = |E_0(\xi,\zeta)|^2 + \mathscr{E}\{|\Delta E(\xi,\zeta)|^2 \}
 \end{equation}
 From \refeqn{eq:deltaE_slot} we therefore find,
 \begin{equation}
 \mathscr{E}\{|\Delta E(\xi,\zeta) |^2 \} = \sigma^2 \sum_{i=1}^N {\left \{ \left [ \int_{-1/2}^{1/2}{f_i(y)e^{-2\pi i \zeta y}dy} \right ]^2 + \left [ \int_{-1/2}^{1/2}{f'_i(y)e^{-2\pi i \zeta y}dy} \right ]^2 \right \} } \le 10^{-10} |E_0(0,0)|^2
 \end{equation}
  
  As we did with the apodization error, we can model each of the mask edge error functions, $f_i(y)$, via a harmonic  expansion.  Since the small amplitude of the error was absorbed into the $\epsilon_i$, the terms in brackets are bounded by one at the  worst locations in the image plane.  Thus, a conservative bound on the average edge error is given by:
  \begin{equation}
  \sigma \le \frac{10^{-5}|E_0(0,0)|}{\sqrt{2N}}
  \end{equation}   
  
For the barcode mask  shown in Section~\ref{sec:one_d}, which has 44 edges per side, this translates into a manufacturing accuracy on each edge of the mask of approximately  $5 \times 10^{-7}$.  For a 2 inch mask, this corresponds to a 25 nm accuracy  requirement on the mask edges, something achievable with current fabrication technologies.

\section{Final Remarks}

In this paper we presented all the asymmetric shaped
pupil masks we have studied to date for achieving high contrast imaging.  All of these masks use apodization at a single pupil to shape the point spread function of the imaging system, achieving $10^{-10}$ contrast outside some desired inner working angle.  These masks can be used in either rectangular or circular apertures, depending upon the specific mission tradeoffs.  While the idea of using pupil apodization has been around for many years, the problems associated with manufacturing such smoothly varying masks to the accuracy needed for planet finding has yet to be solved.  We show here that any apodized pupil can be made into a shaped pupil mask at a small cost in discovery space but with enormous practical benefit.  Shaped pupils are far easier and less expensive to manufacture, are more tolerant to errors, and, in some cases, provide more throughput.  While classical Lyot coronagraphs typically have more throughput, they suffer from the same manufacturing difficulties and are far more sensitive to alignments and pointing error. 

\section*{Acknowledgements}
We gratefully acknowledge the support of the National Aeronautics and Space Administration through the Jet Propulsion Laboratory, California Institute of Technology for this work.


\end{document}